\documentclass[letterpaper,draft,onecolumn,11pt]{IEEEtran}

\renewcommand{\baselinestretch}{1.6}
\usepackage{amsmath,amssymb,bm}
\usepackage{epsfig,subfigure,colortbl,psfrag,subfloat}
\usepackage{enumerate,array,multirow}
\usepackage{cite,url}
\usepackage[subnum]{cases}
\usepackage{threeparttable,supertabular}

\def \beq { \begin{equation} }
\def \eeq { \end{equation} }
\def \beqn{ \begin{eqnarray} }
\def \eeqn{ \end{eqnarray} }
\def \bmat{ \begin{bmatrix} }
\def \emat{ \end{bmatrix} }
\def \bmats{ \left[ \begin{smallmatrix} }
\def \emats{ \end{smallmatrix} \right]}
\def \beqi{\begin{IEEEeqnarray}{rcl}\IEEEyesnumber}
\def \eeqi{\end{IEEEeqnarray}}

\def \minj{\operatornamewithlimits{min}}
\def \maxj{\operatornamewithlimits{max}}

\def \max{\operatornamewithlimits{max}}
\def \ln{\operatorname{ln}}
\def \log{\operatorname{log}}
\def \exp{\operatorname{exp}}

\def \E{\operatorname{E}}

\def \dB{\operatorname{dB}}
\def \dBm{\operatorname{dBm}}

\def \W{\operatorname{W}}
\def \m{\operatorname{m}}
\def \ms{\operatorname{ms}}
\def \us{\operatorname{\mu s}}

\def \bpsHz{\operatorname{b/s/Hz}}
\def \bps{\operatorname{b/s}}

\def \MbpJ{\operatorname{Mb/J}}

\def \MHz{\operatorname{MHz}}
\def \Hz{\operatorname{Hz}}
\def \GHz{\operatorname{GHz}}

\def \Pr{\operatorname{Pr}}

\newcommand{\w}{w}
\newcommand{\x}{x}
\newcommand{\X}{\widetilde{x}}
\newcommand{\y}{y}
\newcommand{\Y}{\widetilde{y}}
\newcommand{\Yvec}{\widetilde{\bm y}}
\newcommand{\yvec}{\bm y}
\newcommand{\Xvec}{\widetilde{\bm x}}
\newcommand{\xvec}{\bm{x}}
\newcommand{\Pmaxin}{{P}^{\rm{max}}_{\rm{in}}}
\newcommand{\Pmaxout}{{P}^{\rm{max}}_{\rm{out}}}
\newcommand{\Pin}{{P}_{\rm{in}}}
\newcommand{\Pout}{{P}_{\rm{out}}}
\newcommand{\Pc}{{P}_{\rm{c}}}
\newcommand{\fPA}{{\sf{L}}_{\rm PA}}
\newcommand{\conv}{\ast}
\newcommand{\circconv}{\otimes}

\newcommand{\yreal}{y_{\rm Re}}

\newcommand{\rhomax}{\rho_{\rm max}}
\newcommand{\bw}{\Omega}
\newcommand{\ampa}{a}
\newcommand{\ampb}{b}
\newcommand{\ampr}{r}
\newcommand{\amax}{a_{\rm max}}
\newcommand{\bmax}{b_{\rm max}}
\newcommand{\ampA}{A}

\newcommand{\awgn}{z}
\newcommand{\awgN}{Z}
\newcommand{\way}{\ell}
\newcommand{\FN}{K}
\newcommand{\Find}{k}

\newcommand{\gsigmaxsquare}{g\Pin}
\newcommand{\gPin}{g\Pin}
\newcommand{\Ppa}{P_{\rm PA}}
\newcommand{\Ncp}{N_{\sf{CP}}} 

\newtheorem{thm}{Theorem}

\newtheorem{prop}[thm]{Proposition}

\psfull

\begin{document}

\pagestyle{empty}

\title{Spectral Efficiency and Energy Efficiency of OFDM Systems: Impact of Power Amplifiers and Countermeasures}
\author{Jingon~Joung,~
        Chin~Keong~Ho,~
        and~Sumei~Sun
%
\thanks{Parts of this work have been presented at the {\it IEEE Global Telecommunications Conference (GLOBECOM)}, Anaheim, CA, USA, Dec. 2012 \cite{JoHoSu12GC}, and the {\it Asia Pacific Signal and Information Processing Association (APSIPA)}, Annual Summit and Conference, Hollywood, CA, USA, Dec. 2012 \cite{JoHoSu12APSIPA}.}
\thanks{The authors are with the Institute for Infocomm Research (I$^2$R), A$^\star$STAR, Singapore 138632 (e-mail: \{jgjoung, hock, sunsm\}@i2r.a-star.edu.sg)}
}
%
\maketitle
\pagestyle{empty}
\thispagestyle{empty}

\vspace{-1cm}
\begin{abstract}
In wireless communication systems, the nonlinear effect and inefficiency of power amplifier (PA) have posed practical challenges for system designs to achieve high spectral efficiency (SE) and energy efficiency (EE). In this paper, we analyze the impact of PA on the SE-EE tradeoff of orthogonal frequency division multiplex (OFDM) systems. An ideal PA that is always linear and incurs no additional power consumption can be shown to yield a decreasing convex function in the SE-EE tradeoff. In contrast, we show that a practical PA has an SE-EE tradeoff that has a turning point and decreases sharply after its maximum EE point. In other words, the Pareto-optimal tradeoff boundary of the SE-EE curve is very narrow. A wide range of SE-EE tradeoff, however, is desired for future wireless communications that have dynamic demand depending on the traffic loads, channel conditions, and system applications, e.g., high-SE-with-low-EE for rate-limited systems and high-EE-with-low-SE for energy-limited systems. For the SE-EE tradeoff improvement, we propose a PA switching (PAS) technique. In a PAS transmitter, one or more PAs are switched on intermittently to maximize the EE and deliver an overall required SE. As a consequence, a high EE over a wide range SE can be achieved, which is verified by numerical evaluations: with $15\%$ SE reduction for low SE demand, the PAS between a low power PA and a high power PA can improve EE by $323\%$, while a single high power PA transmitter improves EE by only $68\%$.
\end{abstract}


\section{Introduction}\label{Sec:Intro}
\begin{figure}[!t]
\begin{center}
\psfrag{j}[cc][cb][.75][0]{$\Ppa\dBm$}
\psfrag{i}[cc][ct][.75][0]{$\Pmaxout\dBm$}
\psfrag{v                                                 }[lc][cc][.7][0]{{\sf PAs from data sheets}}
\psfrag{o}[lc][cc][.7][0]{{\sf 20\% drain efficiency line}}
\psfrag{r}[lc][cc][.7][0]{{\sf 30\% drain efficiency line}}
\psfrag{n}[lc][cc][.7][0]{{\sf 100\% drain efficiency line}}
\psfrag{P}[cc][cc][.8][90]{$\Pout$}
\psfrag{Q}[cc][cc][.8][0]{$\Pin$}
\psfrag{a}[cr][cc][.8][0]{$(\Pmaxout,\Pmaxin)$}
\psfrag{1}[cc][cc][.7][0]{$1\dB$}
\psfrag{s}[cc][cc][.7][0]{\sf saturation}
\psfrag{z}[lc][cc][.7][0]{\sf $1\dB$ compression point}
\psfrag{c}[cc][rt][.7][0]{0}
\psfrag{e}[lc][cc][.7][0]{\sf soft limiter model}
\psfrag{h}[lc][cc][.7][0]{\sf Rapp model}
\psfrag{9}[lc][cc][.7][0]{\sf perfectly linear model}
\subfigure[\!\!\!\!\!\!\!\!\!\!\!\!]{
\includegraphics[height=50mm,width=100mm]{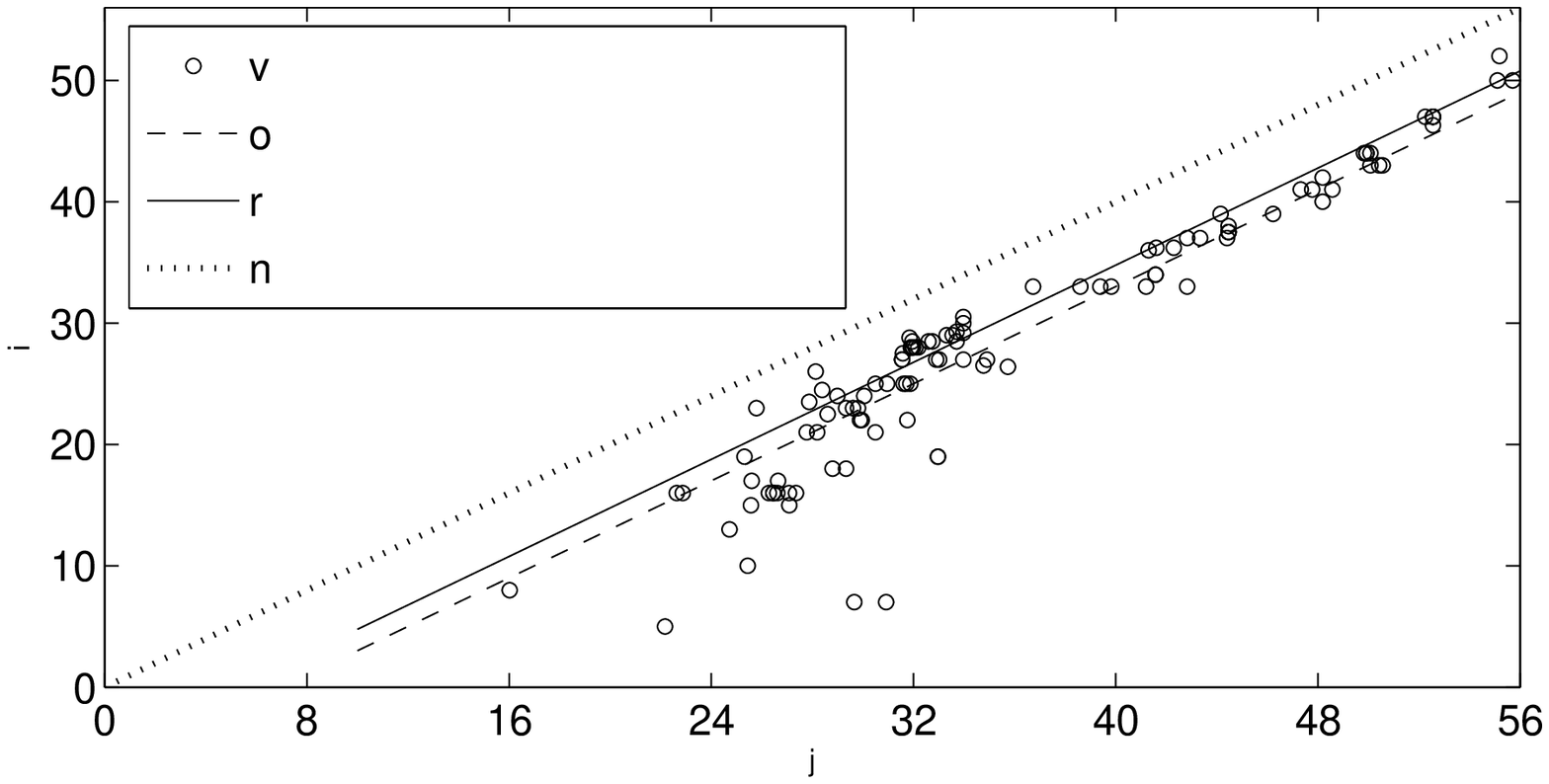}}\\
\subfigure[\!\!\!\!\!\!\!\!\!\!\!\!]{
\includegraphics[height=50mm,width=100mm]{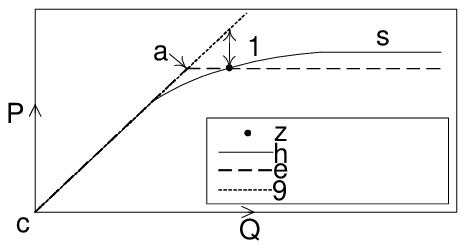}}
\caption{Two fundamental characteristics of a practical PA. (a) Efficiency: maximum output power $\Pmaxout$ (at the linear region) versus consumed power $\Ppa$. (b) Nonlinearity: PA output power $\Pout$ versus PA input power $\Pin$.}\label{Fig:FundPAChar}
\end{center}
\end{figure}

Wireless access communication networks consume significant amount of energy to overcome fading and interference, compared to fixed line communication networks \cite{VeHeDePuLaJoCoMaPi11,BaAyHiTu11}. In wireless networks, energy is mostly consumed at the base station (BS) \cite{VeHeDePuLaJoCoMaPi11}, of which a substantial fraction of $50\%$--$80\%$ of overall power is consumed at power amplifiers (PAs) \cite{BoCo11}. A measure of the PA efficiency is given by the drain efficiency $\eta$ that is the ratio of PA output power $\Pout$ to PA power consumption $\Ppa$, i.e., $\eta=\Pout / \Ppa$. Fig. \ref{Fig:FundPAChar}(a) plots PA maximum output power $\Pmaxout$ versus $\Ppa$, based on our survey of commercially available PAs for which we give a summary of the key parameters in Table \ref{Table:DataSheet} in Appendix \ref{Appendix:Table}. From Fig. \ref{Fig:FundPAChar}(a), we see that $\eta$ at $\Pmaxout$ is typically between $20\%$ and $30\%$, which confirms that the overhead incurred at PA is substantial. To ensure high energy efficiency (EE), the PA characteristics have to be carefully considered in system designs.

On the other hand, high spectral efficiency (SE) is needed to support the growing demands of high-rate applications. Orthogonal frequency division multiplex (OFDM) and orthogonal frequency division multiple access (OFDMA) are two popular spectral efficient systems. However, OFDM and OFDMA modulated signals exhibit high peak-to-average power ratio (PAPR), thus suffering from severe nonlinearity effects \cite{LiCi98, BaSiGoSa02} as illustrated in Fig. \ref{Fig:FundPAChar}(b), in which PA output power $\Pout$ is shown over the PA input power $\Pin$. Two commonly used models, namely the Rapp model \cite{Rapp91} and the soft limiter model \cite{TeHoCi03}, are shown in Fig. \ref{Fig:FundPAChar}(b). They are used to describe the nonlinear amplitude (i.e., signal power) distortion, especially at a high power region, while the phase is assumed to be undistorted (the details will be given in Section \ref{Sec:SysModel}, and for more nonlinearity models, refer to the references in \cite{JoHoSu12APSIPA}). In practice, to circumvent the resulting performance degradation, input backoff (IBO) is implemented by reducing the power of the input signal at the PA, so that the amplification stays within the linearity region as much as possible. While IBO allows high SE to be achieved, it can reduce the EE, because the PA efficiency is typically designed to peak near the saturation point and it usually drops rapidly as the input power decreases \cite{ShSuWo02}. Hence, a tradeoff between SE and EE is inevitable while optimizing with respect to the PA. It is thus important to jointly characterize the role that a PA plays in both SE and EE of wireless communication systems. Recently, circuit power consumption has been taken into consideration for energy efficient system designs \cite{RiFeFe09,IsFeDr10,MiHiLiTa12}, but without consideration of the nonlinearity of the PA.

In this paper, the tradeoff of SE and EE for OFDM systems is analyzed by taking into account the impact of practical PAs that is both {\it inefficient} and {\it nonlinear}. To provide tractable results, we assume that the nonlinearity of the PA is modeled by a soft limiter. To capture the PA inefficiency, we propose a nonlinear transmit power model depending on the PA types. We further provide theoretical results to achieve maximum SE and maximum EE from our analysis, and verify the theoretical results through simulations using real-life device parameters. Consequently, it is shown that the practical SE-EE tradeoff increases before a turning point and decreases rapidly after the turning point. In other words, the PA can support a narrow SE-EE tradeoff with only a limited range of SE. In cellular communications, however, a wide range of SE-EE tradeoff is desired because the BSs need high data rates intermittently, yet need to save energy whenever possible to save operation costs. To achieve a wide Pareto-optimal SE-EE tradeoff region, we propose a PA switching (PAS) technique, in which one or more PAs are switched on at any time to maximize the EE while satisfying the required SE, resulting in a high EE over a wide SE range. For example, with $15\%$ SE reduction for low SE demand, the PAS between a low power PA ($25\W$ maximum power) and a high power PA ($100\W$ maximum power) can improve EE by $323\%$, while a single high power PA transmitter improves EE by only $68\%$. Specifically, our key contributions are summarized as follows:
\begin{itemize}
\item {\it Practical SE}: We obtain a closed-form expression of SE with consideration of PA nonlinearity, and show that its approximation is a concave function with a unique maximum with respect to the input power of the PA.
\item {\it Practical EE}: We establish a PA-dependent nonlinear power consumption model from various recent studies on empirical power measurement and parameters for cellular and wireless local area networks. We show that the EE is a piecewise quasi-concave function with a unique maximum point if the PA is perfectly linear.
\item {\it PAS}: We observe that the practical SE-EE tradeoff decreases rapidly after a turning point, i.e., the limited SE-EE tradeoff for dynamic traffic conditions. To circumvent this, we propose a PAS technique. Numerical results show that the SE-EE tradeoff improvement is significant even though practical losses are considered, such as switch insertion loss and switching time overhead.
\end{itemize}

\section{Prologue}
This paper attempts to quantify analytically and numerically the degradation of both SE and EE caused by the practical nonlinearities and energy consumption of the PA. Specifically, we define SE, in $\operatorname{b/s/Hz}$, as the amount of bits that are reliably decoded per channel use (i.e., per unit time and per unit bandwidth). We define EE, in $\operatorname{b/J}$, as the total amount of reliably decoded bits normalized by the energy. Thus, SE and EE are given respectively by \cite{Verdu02,ChZhXuLi11}
\beqi\label{Tradeoff}
{\sf SE} &~=~& \frac{I(\widetilde{\bm X} ;\widetilde{\bm Y})}{N} \IEEEyessubnumber\label{SE0}\\
{\sf EE} &~=~& \frac{T\bw \; {\sf SE}}{T\Pc} = \frac{\bw \; {\sf SE}}{\Pc} \IEEEyessubnumber\label{EE0}.
\eeqi
Here, $I(\widetilde{\bm X} ; \widetilde{\bm Y})$ is the mutual information in $\operatorname{b/s/Hz}$ given the length-$N$ transmitted and received vectors $\widetilde{\bm X}$ and $\widetilde{\bm Y}$, representing an achievable sum rate over $N$ channel uses \cite{CoTh06book}; $\bw$ is the total bandwidth used; $T$ is the total time used; and $\Pc$ is the total power consumption including the PA power consumption $\Ppa$.

\begin{figure}[!t]
\psfrag{a}[cc][cc][.9][0]{\sf $\widetilde{\bm x}$}
\psfrag{e}[cc][cc][.9][0]{$\widetilde{\bm y}$}
\psfrag{F}[cc][cc][.8][90]{\sf IDFT ${\bm F}$}
\psfrag{H}[cc][cc][.65][90]{\sf add CP $\rightarrow$ P/S $\rightarrow$ DAC}
\psfrag{G}[cc][cc][.65][90]{\sf ADC$\rightarrow$S/P$\rightarrow$remove CP}
\psfrag{K}[cc][cc][.8][90]{\sf DFT ${\bm F}^H$}
\psfrag{z}[cc][cc][.7][0]{${\bm F}$}
\psfrag{m}[cc][cc][.7][0]{${\bm F}^H$}
\psfrag{x}[cc][cc][.9][0]{${\bm x}$}
\psfrag{P}[cc][cc][.9][0]{\sf PA}
\psfrag{o}[cc][cc][.8][0]{$\fPA(\cdot)$}
\psfrag{A}[cc][cc][.7][0]{$\{h_0,\cdots,h_{L-1}\}$}
\psfrag{w}[cc][cc][.9][0]{${\bm w}$}
\psfrag{n}[cc][cc][.9][0]{${\bm \awgn}$}
\psfrag{y}[cc][cc][.9][0]{${\bm y}$}
\begin{center}
\epsfxsize=0.7\textwidth \leavevmode \epsffile{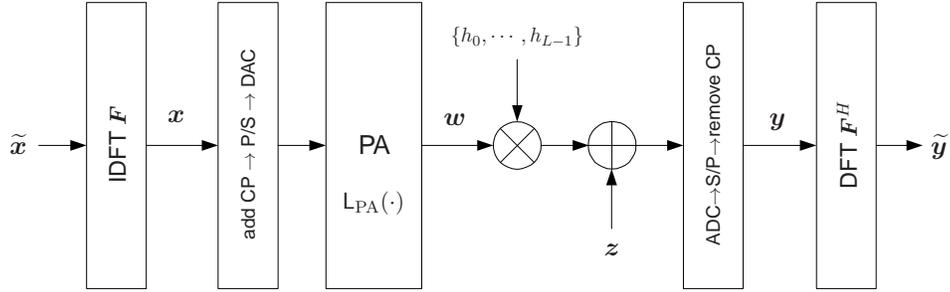}
\caption{An OFDM system with a nonlinear memoryless PA represented by function $\fPA(\cdot)$, assuming perfect synchronization.}
\label{Fig:SystemModel}
\end{center}
\end{figure}

For illustration, consider an {\em ideal} system without system overhead power consumption, i.e., $\Pc=\Ppa$. Furthermore, consider the {\em ideal PA} which is always {\em perfectly efficient} (dotted line in Fig. \ref{Fig:FundPAChar}(a)), i.e,. $\Ppa=\Pout$, and always {\em perfectly linear} (dotted line in Fig. \ref{Fig:FundPAChar}(b)). Using Gaussian signalling, which is optimal for the ideal PA, we get ${\sf SE}=\log_2(1+\Pout/\sigma^2)$, where $\sigma^2$ is the noise power \cite{CoTh06book}. For the ideal system with ideal PA, therefore, asymptotically as $\Pc$ increases, ${\sf SE}$ increases proportionally with $\log_2(\Pc)$ and hence ${\sf EE}$ decreases proportionally with $\log_2(\Pc)/\Pc$. In other words, the SE-EE tradeoff region is a decreasing convex as observed in \cite{Verdu02,ChZhXuLi11} when the system and PA are ideal. In contrast, for a practical system, asymptotically as $\Pc$ increases, the output saturates and so ${\sf SE}$ saturates to some upper limit, hence EE decreases proportionally with $1/\Pc$; moreover, significant overhead power exists, because $\Pc>\Ppa>\Pout$. To account for the degradation of both SE and EE in practice, it is essential to consider practical overhead in system power consumption $\Pc$ and have a sufficiently accurate, yet tractable, model for the PA (i) on its energy consumption to specify the relationship of $\Ppa$ and $\Pout$, and (ii) on its nonlinearity behavior. In the sequel, we shall address both issues when we determine the SE and EE in (\ref{Tradeoff}).

\section{System Model}\label{Sec:SysModel}

Consider the OFDM system with a nonlinear memoryless PA shown in Fig.~\ref{Fig:SystemModel}. Without loss of generality (w.l.o.g.), we consider the transmission of one OFDM symbol which consists of $N$ complex-valued data symbols, denoted by the data vector 
${\Xvec}=\left[\X_0,\cdots ,\X_{N-1}\right]^T$.
The data symbol $\X_n$ is sent on the $n$th orthogonal subcarrier. These $N$ subcarriers occupy a total frequency band of $\bw\Hz$.
The data symbols are assumed to be identical and independently distributed (i.i.d.) subject to the power constraint $\E[|\X|^2]\leq \Pin$;
here and subsequently, we drop the subcarrier or time index if there is no dependence on it or  when there is no ambiguity.
We transform $\Xvec$ to the time domain signal vector
${\xvec}= \left[\x_0 ,\cdots,  \x_{N-1} \right]^T$
according to ${\xvec}={\bm F}{\Xvec}$,
where ${\bm F}$ is an $N$-by-$N$ unitary inverse discrete Fourier transform (IDFT) matrix. Thus, $\E[|\x|^2]\leq \Pin$.
Then a cyclic prefix (CP) of length $\Ncp$ is added to $\xvec$ and passed to a parallel-to-serial (P/S) converter, followed by a digital-to-analogue converter (DAC). We assume the DAC, the analogue-to-digital converter (ADC) and subsequent processing (such as timing and frequency synchronization) are ideal such that, w.l.o.g., we use $\x_t$ to represent the output of the DAC at discrete time index $t$. We rewrite $\x_t=\ampa_t e^{j\theta_t}$ where $\ampa_t\triangleq |\x_t|$ is the amplitude and $\theta_t$ is the phase of $x_t$ where $0\leq\theta_t<2\pi$.

Next, the DAC output $x_t$ is amplified through a memoryless PA described by a nonlinear function $\fPA(\cdot)$ to give the output $\w_t=\fPA(x_t)$, denoted collectively by the vector ${\bm w}=[\w_0,\cdots,\w_{N+\Ncp-1}]^T$. Under the Rapp and soft limiter models illustrated in Fig. \ref{Fig:FundPAChar}(b), we can write $\w_t={\ampb}_t e^{j\theta_t}$ where ${\ampb}_t\triangleq |\w_t|$ while the phase remains the same as that of $\x_t$. Specifically, the Rapp model describes the amplitude distortion according to
\beq\nonumber
\fPA(\ampa_t) = \sqrt{g} \ampa_t \left( 1+\left(\frac{\sqrt{g} \ampa_t}{b_{\rm sat}}\right)^{2p}\right)^{-\frac{1}{2p}},
\eeq
where $\sqrt{g}\geq 1$ is a parameter interpreted as the desired linear gain; $b_{\rm sat}$ is the saturation amplitude when $\ampa_t\rightarrow \infty$; and $p$ controls the smoothness of the transition from the linear region to the saturation region. Thus, the gain is nonlinear for all input signals. For the soft limiter model, the amplitude distortion follows
\beq\nonumber
\fPA(\ampa_t)=
\begin{cases}
\sqrt{g}\ampa_t , & {\text{if }} \ampa_t <  \amax \\
\bmax , & {\text{if }}\ampa_t \geq \amax,
\end{cases}
\eeq
where $\amax \triangleq \sqrt{\Pmaxin}$ and $\bmax\triangleq\sqrt{\Pmaxout}$.
Thus, the output of soft limiter is clipped to a constant $\bmax$ if the input signal exceeds a threshold value $\amax$, and experiences a linear scaling of its input with gain $\sqrt{g}$ otherwise.

Finally, the PA output is transmitted through an $L$-tap multipath channel $\{h_0,h_1,\cdots,h_{L-1} \}$. Assuming $L\leq \Ncp$ and perfect timing synchronization, the CP is removed and the received signal is given by
\beq\label{RxSig}
\y_t=h_t \circconv  \w_t +\awgn_t \triangleq \ampr_t e^{j\phi},\\
\eeq
for $t=0,\cdots, N-1$ (for convenience, we shift the time indices to start from $0$). Here, $\circconv$ is the circular convolution operator, $\awgn_t\sim\mathcal{CN}(0,\sigma_z^2)$ is an additive white Gaussian noise (AWGN), and $\ampr_t$ and $\phi_t$ represent the amplitude and phase of $\y_t$. The received signal vector ${\yvec} = [\y_0, \cdots, \y_{N-1}]^T$ is transformed via a DFT (i.e., a Hermitian transpose of $\bm F$) to give the frequency domain signal vector $\Yvec=[\Y_0, \cdots, \Y_{N-1}]^T={\bm F}^{H} {\yvec}$.

In practice, the time-domain signal after IDFT typically produces a Gaussian-like signal with a high PAPR. It is well known that the nonlinearity of a PA can thus result in significant degradation of the achievable rate of the signal \cite{LiCi98}.
To analytically model the high PAPR and the nonlinearities, we make the following assumptions:
\begin{itemize}
\item[$A1$:]
We assume that the data symbols are i.i.d. with complex normal distribution with zero mean and $\Pin$ variance, denoted as $\X\sim\mathcal{CN}(0,\Pin)$. Hence the time-domain signals are also i.i.d. with distribution $\x\sim \mathcal{CN}(0, \Pin)$. The time domain signals have very high PAPR and thus they are representatives of the scenario when a high-order modulation is used or when $N$ is large.
\item[$A2$:] For tractability of subsequent analysis, we employ the soft limiter model for the PA.
A good approximation of the maximum power output $\Pmaxout$ is given by the one-dB input compression output, where the output power drops $1\dB$ below the desired power output if the gain is linear as illustrated in Fig. \ref{Fig:FundPAChar}(b). Thus, the maximum power input is $\Pmaxin=\Pmaxout/g$.
We shall use data sheets of commercially available products (e.g., Table \ref{Table:DataSheet} in Appendix \ref{Appendix:Table}) to extract suitable parameters for $g$ and $\Pmaxin$ to obtain numerical results. The soft limiter model is analytically tractable, and it can capture the clipping effect in the high power region as the Rapp model (the Rapp model approaches the soft limiter model as $p$ increases). Nonlinearity in low power region of the soft limiter can be assumed to be mitigated by applying linearization techniques, such as feedforward, feedback, and predistortion (refer to the references in \cite{JoHoSu12APSIPA}), which is the same as the Rapp model.
\end{itemize}
The assumption $A1$ is independent to the assumption of the soft limiter model in $A2$, because the probability density functions (pdfs) of $x$ and $\widetilde{x}$ do not change regardless of the PA model. In this paper, we focus on point-to-point communications. The spectral regrowth arisen from the nonlinearity of the PA, which increases the adjacent channel interferences to neighboring bands, is not considered explicitly.

Typically, an IBO is performed to mitigate the degradation resulting from PA nonlinearities, by reducing the input signal power $\Pin$ such that it is much less than $\Pmaxin$.
To reflect this, we write $\Pin=\xi \Pmaxin$, where $\xi\geq 0$ is a power loading factor and is related to the IBO as ${\rm IBO}\triangleq 10\log_{10}(\xi^{-1})\dB$.
By varying $\xi$, we can then perform IBO to tradeoff between EE and SE.

Based on assumptions $A1$ and $A2$, we shall obtain tractable results which offer insights on how the PA affects the SE and EE in Sections~\ref{Sec:SpectralEfficiency} and \ref{Sec:EnergyEfficiency}, respectively. Then we study how this leads to the analysis of a new PA architecture in Section~\ref{Sec:TimeSharing}, which improves SE and EE tradeoff.

\section{Spectral Efficiency}\label{Sec:SpectralEfficiency}

\renewcommand{\x}{X}
\renewcommand{\X}{\widetilde{X}}
\renewcommand{\y}{Y}
\renewcommand{\Y}{\widetilde{Y}}
\renewcommand{\Yvec}{\widetilde{\bm Y}}
\renewcommand{\yvec}{\bm Y}
\renewcommand{\Xvec}{\widetilde{\bm X}}
\renewcommand{\xvec}{\bm{X}}
\renewcommand{\w}{W}

In this section, we determine the SE in \eqref{SE0} under assumptions $A1$ and $A2$. To this end, we obtain the mutual information $I(\Xvec ;\Yvec)$ for flat fading channels in Section~\ref{sec:flat}, and for multipath channels in Section~\ref{sec:multipath}. For simplicity, we ignore the throughput loss due to the addition of the CP. We fix the following PA-related parameters: the power loading factor $\xi$, the gain $g$ in the linearity region and the maximum power output $\Pmaxout$. Thus the maximum input power $\Pmaxin=g^{-1}\Pmaxout$ is also fixed; for convenience, let $\gamma\triangleq\Pmaxout/\sigma_z^2>0$ be the maximum power output normalized by the noise variance $\sigma_z^2$.

We use upper case letters to represent random variables, such as $X$, $W$, and $Y$, and lower case letters to represent their realizations, such as $x$, $w$, and $y$. The pdf of random variable $X$ is denoted by $f_{X}(\cdot)$. Recall that the signals are written in terms of their amplitudes and phases as $x=\ampa e^{j\theta}$, $w=\ampb e^{j\theta}$, and $y=\ampr e^{j\phi}$.

\subsection{Mutual Information in Flat Fading Channel}\label{sec:flat}
Consider the flat fading channel where the number of multipath is $L=1$. Let $h_0=1$, w.l.o.g., as the actual channel attenuation and any fixed energy losses incurred can be reflected by adjusting the noise variance such that the signal-to-noise ratio (SNR) is maintained. Given input $X=Ae^{j\theta}$, the channel model at time index $t$ is
\beqn\label{flatfading}
\y_t &=& \w_t+ \awgN_t, \;\mbox{ where } \w_t = \fPA(\ampA_t) e^{\theta_t}.  
\eeqn
The SE, which is given by the achievable rate averaged over $N$ transmissions, is
\beqn\label{MI}
I(\Xvec ;\Yvec)/N\!\!\!\!\!\!\!\!\!\!
&&\overset{(a)}{=} I(\xvec ;\yvec)/N \nonumber\\
&&\overset{(b)}{=} \sum_{t=0}^{N-1}I(\x_t;\y_t)/N\nonumber\\
&&\overset{(c)}{=} I(\x;\y)\nonumber\\
&& \overset{(d)}{=} H(\y)-\log_2 \pi e \sigma_z^2~[\bps].
\eeqn
Here, (a) follows from the facts that the frequency-domain signals (transmitted and received vectors $\widetilde{\bm X}$ and $\widetilde{\bm Y}$) and time-domain signals (transmitted and received vectors ${\bm X}$ and ${\bm Y}$) are related by a unitary transform, which does not change the mutual information; (b) follows from the independence of the signals in the time domain (because of the memoryless PA and the i.i.d. transmitted signals and noise); (c) follows from the fact that the mutual information is identical over time, and so the time index can be dropped; and (d) follows from the facts that $I(\x;\y)=H(\y)-H(\y|\x)$, the conditional entropy $H(\y|\x)=H(N)$, and $H(N)$ is the differential entropy of a complex Gaussian random variable with variance $\sigma_z^2$ derived by $\log_2 \pi e \sigma_z^2$. The entropy of $\y$ in (\ref{MI}) is given by \cite{CoTh06book}
\beq\label{Entropy}
\begin{split}
H(Y)& = -\int_y f_Y(y)\log_2 f_Y(y) dy.
\end{split}
\eeq

Nonlinear distortion at the transmitter makes it difficult to derive $f_Y(y)$ in (\ref{Entropy}) directly. To tackle this problem, we define a binary random variable $S$ that denotes whether clipping at the PA occurs, i.e., $S=0$ if $\ampA \leq \amax$ and $S=1$ otherwise, and rewrite the pdf of $y$ as $f_Y(y)= \sum_{i=0,1}f_Y(y,S=i)$. Since $\x=\ampA e^{j\theta} \sim\mathcal{CN}(0,\Pin)$, the random variable $\ampA$ follows the Rayleigh distribution. Thus, we get the probability of $S$ as
\beq\label{PrS}
\begin{split}
\Pr(S=0)&=
\Pr\left(\ampA \leq \amax \right) = 1- \exp\left(-\amax^2 {\Pin^{-1}}\right)\\
&=1- \exp\left(-\xi^{-1}\right)\\
\Pr(S=1)&=1-\Pr(S=0)=\exp\left(-\xi^{-1}\right).
\end{split}
\eeq
The numerical computation of the entropy \eqref{Entropy} is straightforward with a closed-form expression of $f_Y(y,S=0)$ and $f_Y(y,S=1)$, which are derived respectively as follows (see Appendix \ref{Appendix:Eq}):
\beqi\label{fy13}
f_{\y}(y,S=0)
&=& {\sf N}_0(y)\left[1-{\sf Q}_1\left(\sqrt{\mu(y)},\sqrt{\rhomax}  \right)\right]\IEEEyessubnumber\label{fy1}\\
f_{\y}(y,S=1)&~=~&{\sf N}_1(y)\Bigg[ \Pr(S=1)\exp\left(\frac{-2\bmax \yreal }{\sigma_z^2}\right) {\sf I}_0\left(\frac{2\bmax|y|}{\sigma_z^2}\right)\Bigg]\IEEEyessubnumber\label{fy3}
\eeqi
where ${\sf N}_0(y)$ denotes the pdf of $\mathcal{CN}\left(0,\gPin+\sigma_z^2\right)$; ${\sf Q}_1(\cdot,\cdot)$ is the Marcum-Q-function \cite{PaPi02book} with parameters $\rhomax \triangleq \frac{2(\gPin+\sigma_z^2)}{\gPin}\sqrt{\bmax}$ and $\mu(y)\triangleq \frac{8 \gPin (\gPin+\sigma_z^2)}{\sigma_z^4}|y|^2$; ${\sf N}_1(y)$ is the pdf of $\mathcal{CN}\left(\bmax,\sigma_z^2\right)$; $\yreal$ is the real part of $y$; and ${\sf I}_0(\cdot)$ is the modified Bessel function of first kind \cite{PaPi02book}.

\subsection{Mutual Information in Multipath Channel}\label{sec:multipath}

We now consider the general case of an $L$-tap multipath channel, where $1\leq L\leq \Ncp$. The received signal in the time domain is given by \eqref{RxSig}. If the amplification of PA is perfectly linear, then the mutual information is given equivalently in the frequency domain as $I(\xvec ;\yvec)/N= \sum_{k=0}^{N-1} I(\X_k ;\Y_k) /N=\sum_{k=0}^{N-1} \log_2(1+|\widetilde{H}_k|^2 \gPin/\sigma_z^2)/N$, where $\widetilde{H}_k$ is the frequency domain channel, see e.g., \cite{CoTh06book}. In our case of interest, however, the nonlinear PA makes the exact analysis of the mutual information intractable, because the PA nonlinearities result in a correlated interference in the frequency domain which is not formulated as a closed-form expression. Instead, we obtain a lower bound for the mutual information (see Appendix \ref{Appendix:Multi}):
\beqn
I(\Xvec ;\Yvec)/N
&\!\!\!\geq\!\!\!\!& \sum_{t=0}^{L-1}\!I(\x_t;\y_{t},\!\cdots,\y_{t+L-1}| \x_1,\!\cdots, \x_{t-1})/N +\sum_{t=L}^{N-1} I^{\text{LB}}_t /N 
\label{MI_multipath4}
\eeqn
where $I^{\text{LB}}_t$ is the mutual information of flat fading channel (\ref{flatfading}) with the SNR given by the equivalent channel (\ref{MRCch}). As $N\rightarrow \infty$, the first term approaches zero, while the second term equals approaches $I^{\text{LB}}_t$ which is in fact independent of $t$ (we drop the index subsequently). Thus, the lower bound in \eqref{MI_multipath4} is given asymptotically by $I^{\text{LB}}$ for $N\gg L$. Note that $I^{\text{LB}}$ can be computed from (\ref{MI}) directly. Numerical results (not included) show that the bound is typically tight if the power of the multipath decreases exponentially over the channel delay.

\subsection{Analytical Results on SE}
Using (\ref{fy1}) and (\ref{fy3}) into (\ref{Entropy}), we find $H(Y)$ and get $I(X;Y)$ from (\ref{MI}) in flat fading channels. Similarly, from (\ref{MI_multipath3}), we can obtain the mutual information of the signals in multipath channels. Accordingly, we derive the SE in (\ref{SE0}) as a function of $\xi$ as
\beq\label{SE}
{\sf SE}(\xi) = 
H(\y)-\log_2 \pi e \sigma_z^2,
\eeq
where note that the entropy $H(Y)$ is a function of $\xi$ as the conditional probabilities in (\ref{fy1}) and (\ref{fy3}) are functions of $\Pin=\xi\Pmaxin$ and $\bmax=\sqrt{\gsigmaxsquare\xi^{-1}}$.

If the PA is perfectly linear, i.e., $\bmax\rightarrow \infty$ and thus $\amax\rightarrow \infty$, it can be easily checked that $f_{Y}(y,S=0)={\sf N}_0(y)$ from (\ref{rvW})--(\ref{fy}) and $f_{Y}(y,S=1)=0$ from (\ref{fy3a}) as $P(S=0)=1$ and $P(S=1)=0$ in (\ref{PrS}). Thus, $H(Y)=\log_2\pi e\left( g \Pin + \sigma_{\awgn}^2\right)$ and we recover the well-known SE for ideal PA as
\begin{equation}
\begin{split}\label{SEideal}
{\sf SE}^{\rm ideal}(\xi) = 
\log_2\left(1+ \gamma\xi \right).
\end{split}
\end{equation}

\begin{figure}[!t]
\psfrag{x}[cc][cc][.8][0]{$r$}
\psfrag{y}[cc][cc][.8][0]{\sf pdf}
\psfrag{a                                             }[lc][cc][.7][0]{\sf ${\sf N}_0(y_0)$ for $\phi=0$ and $\phi=\pi$}
\psfrag{c}[lc][cc][.7][0]{\sf ${\sf N}_1(y_1)$ for $\phi=0$ and $\phi=\pi$}
\psfrag{u}[lc][cc][.9][0]{$-\bmax$}
\psfrag{v}[rc][cc][.9][0]{$\bmax$}
\begin{center}
\epsfxsize=0.7\textwidth \leavevmode
\epsffile{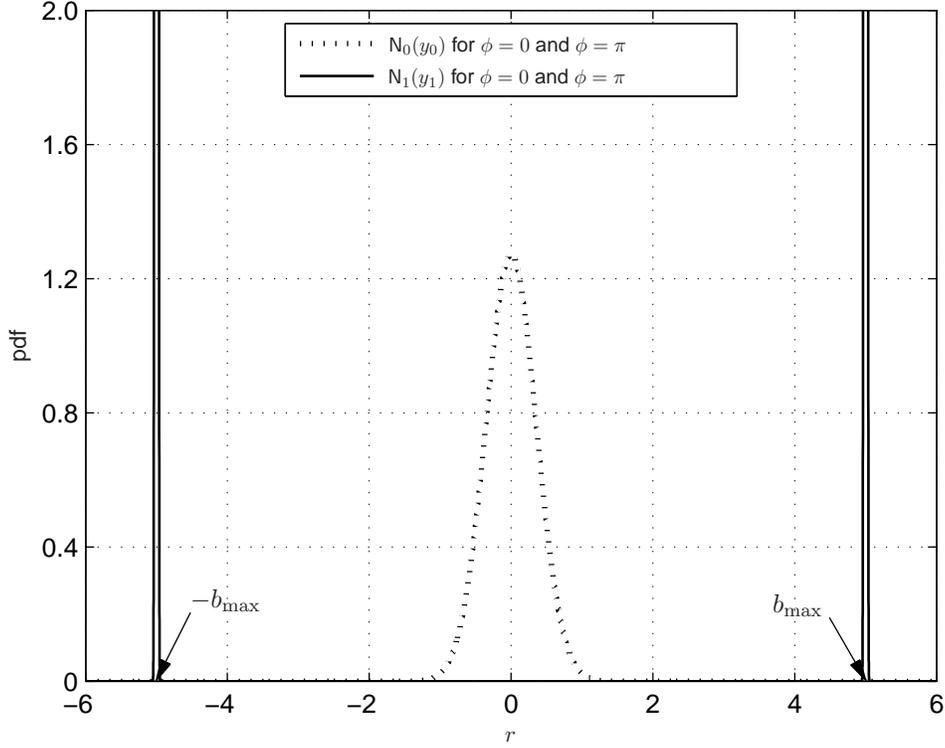}
\caption{The pdfs of ${\sf N}_0(y_0)$ and ${\sf N}_1(y_1)$ for $\phi=0$ and $\phi=\pi$, where $\xi=0.01$. The details for the simulation environment are given in Section \ref{Sec:NumericalSE}.}
\label{Fig:pdfs}
\end{center}
\end{figure}

For tractable analysis, the SE in (\ref{SE}) is approximated under the assumption of low power input signal to PA, i.e., small $\xi$. If $\xi\ll 1$, we can approximate the joint pdfs in (\ref{fy1}) and (\ref{fy3}) as follows:
\beq\label{pdfApprox}
\begin{split}
{f_{Y}}(y,S=0)&\approx{\sf N}_0(y)\\
{f_{Y}}(y,S=1)&\approx{\sf N}_1(y)\Pr(S=1).
\end{split}
\eeq
The approximations comes from the observation that $\xi\ll 1$ implies that the received signal $y$ is also around zero with high probability, i.e., $f_{Y}(y)$ is significant only for $|y|\ll 1$. Thus, $\mu(y)\ll 1$, ${\sf Q}_1(\sqrt{\mu(y)},\sqrt{\rhomax})\approx 0$ in (\ref{fy1}), $\exp(\cdot)\approx1$, and ${\sf I}_0(\cdot)\approx 1$ in (\ref{fy3}), which leads to (\ref{pdfApprox}). Thus, we approximate (\ref{Entropy}) as
\beqi
\widetilde{H}(Y)
&~=~&
-\int_y  \left({\sf N}_0(y)+{\sf N}_1(y)\Pr(S=1)\right) \log_2\left({\sf N}_0(y)+{\sf N}_1(y)\Pr(S=1)\right) dy\nonumber\\
&~\approx~&
-\int_{y_0}{\sf N}_0(y_0)\log_2{\sf N}_0(y_0) dy_0-\int_{y_2} {\sf N}_1(y_1)\Pr(S=1)\log_2{\sf N}_1(y_1)\Pr(S=1)dy_1\IEEEyessubnumber\label{approx2}\\
&~=~&
\log_2\pi e\left(1+\gamma\xi\right) - e^{-\frac{1}{\xi}}\log_2e^{-\frac{1}{\xi}}+ e^{-\frac{1}{\xi}}\log_2\pi e \sigma_z^2\IEEEyessubnumber\label{EntropyApp}
\eeqi
where the approximation in (\ref{approx2}) follows from the further observation that the domains of ${\sf N}_0(y_0)$ and ${\sf N}_1(y_1)$ are approximately disjoint as the gap of their mean values is much larger than their variances, i.e., $\bmax \gg \{\gsigmaxsquare+\sigma_z^2,\sigma_z^2\}$. For example, see Fig. \ref{Fig:pdfs} where ${\sf N}_0(y_0)$ and ${\sf N}_1(y_1)$ are shown for $\phi=\{0,\pi\}$. We note that typically this holds if $\xi\ll 1$, when IBO is used. We thus call the resulting SE as ${\sf SE}^{\rm IBO}$ which is obtained by substituting (\ref{EntropyApp}) to (\ref{SE}) as
\beq\label{SEapprox}
{\sf SE}^{\rm IBO}(\xi) = 
\widetilde{H}(\y)-\log_2 \pi e \sigma_z^2.
\eeq
The following theorems for the approximated SE, ${\sf SE}^{\rm IBO}(\xi)$, allow us to obtain insights on the structured properties of the actual SE, ${\sf SE}(\xi)$, at least for $\xi\ll1$. The proofs are given in Appendix \ref{Appendix:Theorem1}.

\begin{thm}\label{thm1}
The approximated SE, ${\sf SE}^{\rm IBO}(\xi)$, is a concave function over $\max\left(0,-\frac{1}{\ln \pi \sigma_\awgn^2}\right)<\xi \leq \frac{1}{2}$.
\end{thm}
Using {\it Theorem \ref{thm1}}, we obtain the SE-aware optimal power loading factor $\xi_{\sf SE}^{\star}$.
\begin{thm}\label{thm2}
The SE-aware optimal power loading factor $\xi_{\sf SE}^{\star}$ which maximizes ${\sf SE}^{\rm IBO}(\xi)$ is obtained by the solution of the following equality:
\beq\label{C1Eq}
\frac{\gamma}{1+\gamma\xi} = e^{-\xi^{-1}}\xi^{-2}\left(-\xi^{-1}+1-\ln \pi e \sigma_z^2\right).
\eeq
\end{thm}
\begin{prop}\label{prop1}
A closed form approximation of $\xi_{\sf SE}^{\star}$ is given by
\beq\label{XiO}
\xi_{\sf SE}^{\star} \approx  \widetilde{\xi}_{\sf SE}^{\star} \triangleq \frac{-1}{{\sf{W}}\left(\frac{1}{\ln \left(\pi e\sigma_z^2 \right)}\right)},
\eeq
where ${\sf W}(\cdot)$ denotes the Lambert W function\footnote{For $q<0$, ${\rm W}(q)$ can take multiple values. We assume ${\sf W}(\cdot)\leq -1$ which is known as the lower branch of ${\sf W}(\cdot)$, so that $\widetilde{\xi}_{\sf SE}^{\star} \leq 1$. This gives a unique value for ${\sf W}(\cdot)$.} that satisfies $q={\sf W}(q) e^{{\sf W}(q)}$ {\rm \cite{ChMo02}}.
\end{prop}

Interestingly, the approximated $\widetilde{\xi}_{\sf SE}^{\star}$ depends only on $\sigma_{\awgn}^2$; intuitively, this is because we assume $\xi\ll 1$. This makes $\widetilde{\xi}_{\sf SE}^{\star}$ independent of other PA parameters. The typical values of IBO are between $8\dB$ and $12\dB$ for large (e.g., macro) and small (e.g., femto) cell base stations, respectively, which include an additional margin for fading channels \cite{AuGiDeGoSkOlImSaGoBlFe11,EARTH}. The numerical results in the subsequent subsection show that the analytical results with $\xi\ll1$ are accurate for $\xi\leq0.3$, i.e., ${\rm IBO}\geq 5\dB$.

\subsection{Numerical Results on SE}\label{Sec:NumericalSE}
To verify the analytical results on SE, we evaluate ${\sf SE}(\xi)$ with respect to the power loading factor $\xi$. The bandwidth is set to $10\MHz$. For simplicity, Rayleigh fading channel is assumed with zero mean and unit variance. A more realistic multipath channels as given in \cite{WaPaYa07} may also be used for verifying the results obtained in Section \ref{sec:multipath}. The channel attenuation is modeled as follows \cite{LTE}: $G-128+10\log_{10}(d^{-\alpha})\dB$ where $G$ includes the transceiver feeder loss and antenna gains; and $d^{-\alpha}$ is the path loss where $d$ is the distance in kilometers between a transmitter and a receiver and $\alpha$ is a path loss exponent. In simulations, we set $G=5\dB$, $\alpha=3.76$, $d=200\m$, and $\sigma_{\awgn}^2=-174\dBm/\Hz$ and use a PA SM2122-44L ($\Pmaxout=44\dBm=25\W$ and $g=55\dB$) in Table \ref{Table:DataSheet}.

\begin{figure}[!t]
\psfrag{x}[cc][cc][.8][0]{Power loading factor, $\xi$}
\psfrag{y}[cc][cc][.8][0]{${\sf SE}(\xi),~\bpsHz$}
\psfrag{a                                                                               }[lc][cc][.68][0]{${\sf SE}^{\rm ideal}(\xi)$}
\psfrag{c}[lc][cc][.68][0]{${\sf SE}(\xi)$}
\psfrag{e}[lc][cc][.68][0]{${\sf SE}^{\rm IBO}(\xi)$}
\psfrag{n}[lc][cc][.68][0]{${\sf SE}(\widetilde{\xi}_{\sf SE}^{\star})$}
\psfrag{o}[lc][cc][.68][0]{:\sf~ ideal SE in (\ref{SEideal}) with an ideal PA}
\psfrag{s}[lc][cc][.68][0]{:\sf~ practical SE in (\ref{SE}) with a practical PA}
\psfrag{u}[lc][cc][.68][0]{:\sf~ approximated SE in (\ref{SEapprox})}
\psfrag{v}[lc][cc][.68][0]{:\sf~ optimal SE from (\ref{XiO})}
\begin{center}
\epsfxsize=0.7\textwidth \leavevmode
\epsffile{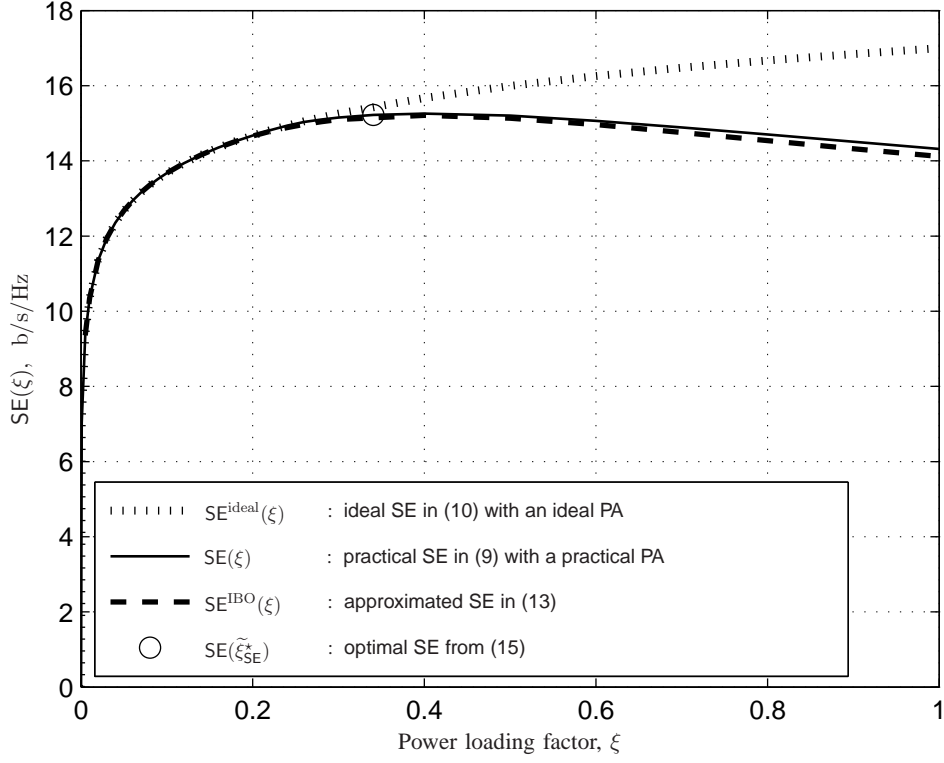}
\caption{Spectral efficiency evaluation with $\Pmaxout=25\W$ and $g=55\dB$.}
\label{Fig:SE}
\end{center}
\end{figure}

Fig. \ref{Fig:SE} shows the numerical evaluation of SE. As expected, ${\sf SE}^{\rm ideal}(\xi)$ in (\ref{SEideal}) achieved by a perfectly linear PA is an increasing concave (log-shape) function, while the practical SE ${\sf SE}(\xi)$ in (\ref{SE}) is a concave function with a unique maximum when $\xi\leq\frac{1}{2}$. The approximated SE ${\sf SE}^{\rm IBO}$ in (\ref{EntropyApp}) matches well with practical SE ${\sf SE}(\xi)$ if $\xi$ is low. The optimal $\widetilde{\xi}_{\sf SE}^{\star}$ in (\ref{XiO}) found from ${\sf SE}^{\rm IBO}(\xi)$ yields almost the highest SE ${\sf SE}(\widetilde{\xi}_{\sf SE}^{\star})$ as marked by `$\circ$' ({\it Theorem \ref{thm2}}). This illustrates the tightness of the approximation made to obtain ${\sf SE}^{\rm IBO}(\xi)$, at least for obtaining the optimal $\xi_{\sf SE}^{\star}$. On the other hand, the discrepancy between the practical SE ${\sf SE}(\xi)$ and the approximated SEs, ${\sf SE}^{\rm ideal}(\xi)$ and ${\sf SE}^{\rm IBO}(\xi)$, increases as $\xi$ (i.e., the PA input or output power) increases.

\section{Energy Efficiency}\label{Sec:EnergyEfficiency}
\begin{figure}[!t]
\psfrag{a}[lc][cc][.9][0]{$P_{\rm BB}$}
\psfrag{b}[lc][cc][.9][0]{$P_{\rm RF}$}
\psfrag{c}[lc][cc][.9][0]{$\Ppa$}
\psfrag{z}[lc][cc][.9][0]{$P_{\rm CB}$}
\psfrag{d}[cc][cc][.9][0]{$\Pin$}
\psfrag{e}[cc][cc][.9][0]{$\Pout$}
\psfrag{m}[lc][cc][.9][0]{\sf loss}
\psfrag{n}[lc][cc][.9][0]{\sf loss}
\psfrag{o}[lc][cc][.9][0]{\sf loss}
\psfrag{u}[lc][cc][.9][0]{\sf loss}
\psfrag{1}[cc][cc][.8][0]{\sf DC}
\psfrag{2}[cc][cc][.8][0]{\sf Power}
\psfrag{3}[cc][cc][.8][0]{\sf Supply}
\psfrag{4}[cc][cc][.8][0]{\sf (PS)}
\psfrag{5}[cc][cc][.8][0]{\sf Module}
\psfrag{6}[cc][cc][.8][0]{\sf Active Cooler and Battery Back up (CB) Module (if present)}
\psfrag{7}[cc][cc][.8][0]{\sf Base Band}
\psfrag{8}[cc][cc][.8][0]{\sf (BB) Module}
\psfrag{9}[cc][cc][.8][0]{\sf Radio}
\psfrag{f}[cc][cc][.8][0]{\sf Frequency}
\psfrag{g}[cc][cc][.8][0]{\sf (RF) Module}
\psfrag{h}[cc][cc][.8][0]{\sf (except PA)}
\psfrag{k}[cc][cc][.8][0]{\sf Power}
\psfrag{p}[cc][cc][.8][0]{\sf Amplifier (PA)}
\psfrag{r}[cc][cc][.8][0]{\sf Module}
\begin{center}
\epsfxsize=0.8\textwidth \leavevmode \epsffile{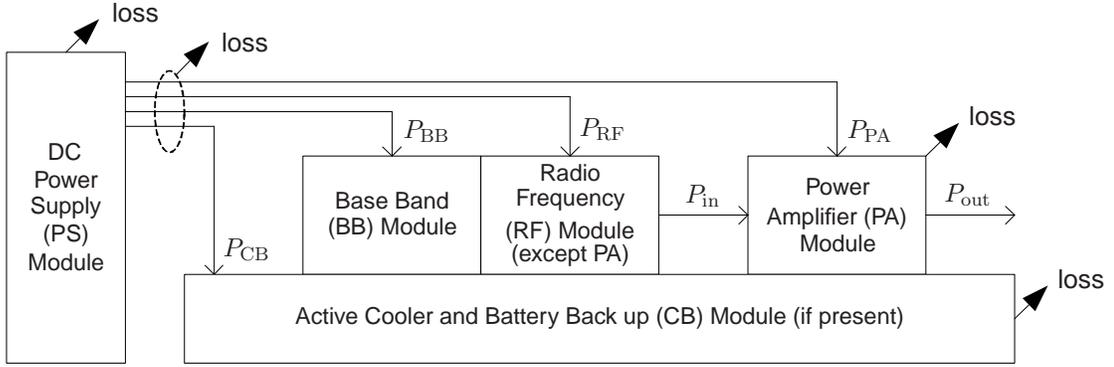}
\caption{Power consumption block diagram including DC power supply (PS), base band (BB), radio frequency (RF), power amplifier (PA), and active cooler and battery back up (CB) modules.}
\label{Fig:PcModel}
\end{center}
\end{figure}

To derive the EE, we first model the power consumption at the transmitter. As shown in Fig. \ref{Fig:PcModel}, power consumption and losses at the transmitter can occur in five modules: a direct current (DC) power supply (PS) module, a base band (BB) module, a radio frequency (RF) module, a PA module, and an active cooler and battery backup (CB) module. Power consumption at BB, RF, PA, and CB modules are denoted by $P_{\rm BB}$, $P_{\rm RF}$, $\Ppa$, and $P_{\rm CB}$, respectively, see details in \cite{AuGiDeGoSkOlImSaGoBlFe11,DeJoMa12,ArRiFeBl11, KuGu11}. After introducing two known power consumption models, we will introduce a new model by taking the PA types (efficiency) and power loading factor $\xi$ into consideration, and subsequently derive the corresponding EE.

\subsection{Existing Power Consumption Models}\label{Sec:PowerConsumptionModel}

One {\it empirical linear} model given in many recent studies, such as \cite{RiFeFe09}, \cite{AuGiDeGoSkOlImSaGoBlFe11}, and \cite{DeJoMa12}, is
\beq\label{ModelI}
\Pc(\xi') =
P_{\rm fix} +   c \xi'\Pmaxout,
\eeq
where $\xi'$ is a frequency loading factor in OFDMA systems ($0 < \xi' \leq 1$); $P_{\rm fix}$ is a power consumption which is independent of the PA output signal power, i.e., $\xi'\Pmaxout$; and $c$ is a scaling coefficient for the power loading dependency. If $\xi'=0$, i.e., at the idle mode, $\Pc(\xi')=P_{\rm idle}$. In Table \ref{Table:Coeffi}, we summarize the parameters $P_{\rm fix}$, $\Pmaxout$, $P_{\rm idle}$, and $c$ for various types of networks. The power coefficient in \cite{DeJoMa12} is modeled as $c=\frac{1}{\eta}+\frac{P_{\rm BB}}{\Pmaxout}+\frac{P_{\rm RF}}{\Pmaxout}$. The parameters depend on the various practical factors, such as the transmitter configuration, the network structure, and the semiconductor technologies employed. For further information, refer to \cite{EARTH}.

       \begin{table}[!t]
        \centering
        \renewcommand{\arraystretch}{1.2}
        \resizebox{.7\textwidth }{!}{
        \begin{threeparttable}
        \caption{Power Model Parameters from \cite{RiFeFe09}$^\dag$, (\cite{EARTH,AuGiDeGoSkOlImSaGoBlFe11})$^\ddag$, \cite{DeJoMa12}$^\S$.}
        \label{Table:Coeffi}
        \begin{tabular}{c | c | c | c | c}\hline
        BS type & $\Pmaxout~\W$  & $P_{\rm fix}~\W$ & $P_{\rm idle}~\W$ & $c$ \\\hline
        Macro & 20$^{\ddag,\S}$ & 130$^\ddag$ 354.44$^\dag$, 405$^\S$ & 75$^\ddag$ & 4.7$^\ddag$, 21.45$^\dag$, 17.8$^\S$\\
        RRH\tnote{$\diamond$}  & 20$^{\ddag}$ & 84$^\ddag$  & 56$^\ddag$ & 2.8$^\ddag$\\
        Micro & 2$^\S$, 6.3$^\ddag$ & 56$^\ddag$, 71.5$^\dag$, 106$^\S$  & 39$^\ddag$ & 2.6$^\ddag$, 7.84$^\dag$, 108.3$^\S$\\
        Pico  & 0.13$^\ddag$ & 6.8$^\ddag$   & 4.3$^\ddag$  & 4$^\ddag$\\
        Femto & 0.05$^\ddag$ & 4.8$^\ddag$   & 2.9$^\ddag$  & 8$^\ddag$\\
        \hline
        \end{tabular}
        {\footnotesize
        \begin{tablenotes}
        \item[$\diamond$] {remote radio head or remote radio unit (RRU)}
        \end{tablenotes}
        }
        \end{threeparttable}
        }
        \end{table}

Since the model in (\ref{ModelI}) is obtained from empirical measurements, it gives a reasonable indication of power consumption; however, no accurate indication is given for the specific PA type used. Furthermore, as shown in \cite{AuGiDeGoSkOlImSaGoBlFe11}, there is a nonlinear relationship between the loading factor and the actual power consumption, especially in high power transmission, e.g., at the macro BS. To address these limitations, a PA-dependent model is given by \cite{ArRiFeBl11,KuGu11}
\beq\label{ModelII1}
\Pc = (1+C_{\rm PS})(1+C_{\rm CB})(P_{\rm BB}+P_{\rm RF}+\Ppa)
\eeq
where $C_{\rm PS}$ is a PS coefficient (typically $0.1\leq C_{\rm PS}\leq0.15$) and $C_{\rm CB}$ is an CB coefficient (typically less than $0.4$). We can modify the model in (\ref{ModelII1}) according to the PA types and the power loading factor $\xi$ because the PA power consumption is modeled explicitly and separately from the other power consumption factors. Note that the frequency loading factor $\xi'$ in (\ref{ModelI}) can be interpreted as the power loading factor $\xi$ in the time domain.

\subsection{Proposed PA-dependant Nonlinear Power Consumption Model}\label{sec:PropPC}
Though the PA power consumption $\Ppa$ depends on many factors including the specific hardware implementation, DC bias condition, load characteristics, operating frequency and PA output power, the component that consumes the majority of the power is given by the DC power fed to the PA \cite{RaAsCrKePoPoSeSo02}. Since the drain efficiency $\eta$ depends on the PA types, we can express $\Ppa$ for different types of PA as a function of $\xi$ \cite{Raab87}. For the $\way$-way Doherty PA, where $\way$ is a fixed positive integer that depends on the implementation, the PA power consumption is expressed as 
\beq\label{PAmodel}
\Ppa(\xi)=\frac{4 \Pmaxout}{\way\pi}\times
\begin{cases}
\sqrt{\xi}, &  0< \xi \leq \frac{1}{\way^2} \\
(\way+1) \sqrt{\xi} - 1, & \frac{1}{\way^2} < \xi \leq1.
\end{cases}
\eeq

Henceforth, we assume the use of the $\way$-way Doherty PA which has widespread use \cite{KiKiMo10,BePrHu11}. The Doherty PA includes the special case of the class B PA with $\way=1$. The PA modeled in (\ref{PAmodel}) can be considered to be an one-stage PA, which is relevant typically for low power transmission. We can obtain $\Ppa(\xi)$ similarly for other PA types, e.g., multi-stage PA combining class-A and Doherty, for high power transmission. It is straightforward to generalize to a multi-stage PA, in which the PA efficiency will change and (\ref{PAmodel}) will be slightly modified accordingly with more levels. However, the EE analysis in the paper will remain without changes in the low power region.

\begin{figure}[!t]
\psfrag{x}[cc][cc][.75][0]{\sf Power loading factor, $\xi$}
\psfrag{y}[cc][cc][.75][0]{\sf Power consumption, $\Pc(\xi),\W$ }
\psfrag{a                                                                              }[lc][cc][.68][0]{\sf idle mode in (\ref{ModelI})}
\psfrag{c}[lc][cc][.68][0]{\sf linear model in (\ref{ModelI})}
\psfrag{e}[lc][cc][.68][0]{\sf nonlinear model with class A PA in (\ref{ModelII2})}
\psfrag{s}[lc][cc][.68][0]{\sf nonlinear model with class B PA in (\ref{ModelII2})}
\psfrag{u}[lc][cc][.68][0]{\sf nonlinear model with $2$-way Doherty PA in (\ref{ModelII2})}
\psfrag{v}[lc][cc][.68][0]{\sf model with ideal PA in (\ref{ModelII3})}
\begin{center}
\epsfxsize=0.7\textwidth \leavevmode \epsffile{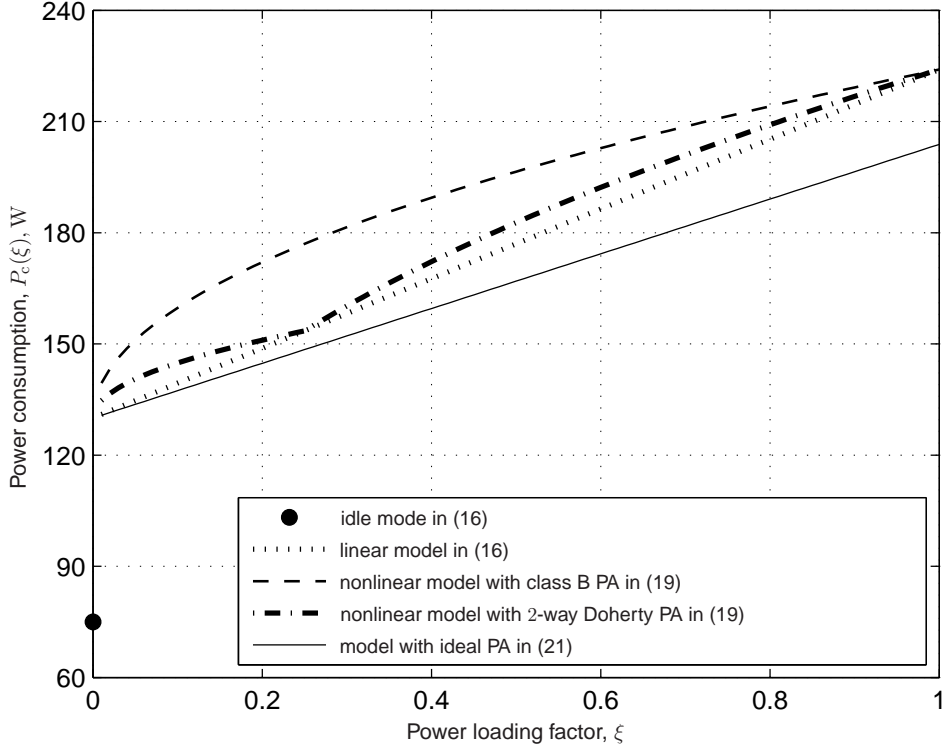}
\caption{Power consumption of microcell BS with $P_{\rm fix}=130\W$ and $c=4.7$.}
\label{Fig:Pc2}
\end{center}
\end{figure}

Substituting (\ref{PAmodel}) to (\ref{ModelII1}), we get a {\it PA-dependant nonlinear} power consumption model as
\beq\label{ModelII2}
\Pc(\xi) = P_{0} + c_0 \left(c_1+c_2\sqrt{\xi}\right) \Pmaxout
\eeq
for $0<\xi\leq 1$, where $P_0=(1+C_{\rm PS})(1+C_{\rm CB})(P_{\rm BB}+P_{\rm RF})$, and
\begin{subnumcases}{\label{Coeffi} (c_1,c_2)=\frac{4}{\way\pi} \times }
\left(0,1\right),  & $0<\xi \leq \frac{1}{\way^2}$,
\label{Coeffi1} \\
\left(-1,\way+1\right),  & $\frac{1}{\way^2} < \xi \leq 1$. \label{Coeffi2}
\end{subnumcases}
Comparing (\ref{ModelII2}) with the model in (\ref{ModelI}), we also see that the new model in (\ref{ModelII2}) reflects the PAs' characteristics. However, since $P_{\rm RF}$ is actually related to $\xi$, there are degrees of freedom to determine $P_0$ and $c_0$. In this work, we set $P_0=P_{\rm fix}$ and $c_0=\frac{\pi}{4}c$, so that (\ref{ModelII2}) matches to (\ref{ModelI}) when $\xi=1$. In other words, this alignment allows us to match the power consumption in (\ref{ModelII2}) with that of (\ref{ModelI}) at the critical points of $\xi$, namely, $\xi=0$ (idle), $\xi=\frac{1}{\way^2}$, and $\xi=1$ as shown in Fig. \ref{Fig:Pc2}. In Fig. \ref{Fig:Pc2}, we use a macrocell setup in Table \ref{Table:Coeffi} where $P_{\rm fix}=130\W$ and $c=4.7$. Following the same procedure of modeling in this subsection, any PA can be reflected in (\ref{ModelII2}).

If a PA is ideal, namely, the PA is perfectly linear and efficient\footnote{Power-added efficiency (PAE) and overall efficiency are defined as $\frac{\Pout-\Pin}{\Ppa}$ and $\frac{\Pout}{\Pin+\Ppa}$, respectively \cite{RaAsCrKePoPoSeSo02}.}, then $\Pout=g\Pin$ and $\Ppa=\Pout-\Pin$, respectively. Thus, $\Ppa=(1-g^{-1})\xi\Pmaxout$. From (\ref{ModelII1}), we can model the PA power consumption with the ideal PA as follows ($0<\xi\leq 1$):
\beq\label{ModelII3}
\Pc^{\rm ideal}(\xi) = P_{\rm fix} + c_0\left(1-g^{-1}\right)\xi \Pmaxout.
\eeq
From Fig.~\ref{Fig:Pc2}, $\Pc^{\rm ideal}(\xi)$ gives a lower bound for the power consumption of the other models, as expected.

\subsection{Analytical Results on EE}
Using the practical SE in (\ref{SE}) and PA-dependent nonlinear power consumption $\Pc(\xi)$ in (\ref{ModelII2}), we obtain the {\em practical} EE given by (\ref{EE0}) as
\begin{equation}\label{EE}
\begin{split}
{\sf EE}(\xi)=\frac{ \bw {\sf SE}(\xi)}{\Pc(\xi)}.
\end{split}
\end{equation}
An upper bound of ${\sf EE}(\xi)$ is obtained assuming an ideal PA with perfect linearity and efficiency as
\beq\label{EEideal}
{\sf EE}^{\rm ideal}(\xi) \triangleq \frac{\bw {\sf SE}^{\rm ideal}(\xi)}{\Pc^{\rm ideal}(\xi)}.
\eeq
However, the bound ${\sf EE}^{\rm ideal}(\xi)$ is not tight enough. Furthermore it does not reflect the PA types. Thus, we remove the perfect efficiency assumption from (\ref{EEideal}) and get a PA-dependant tighter bound as follows:
\beq\label{EELinear}
{\sf EE}^{\rm linear}(\xi) \triangleq  \frac{\bw {\sf SE}^{\rm ideal}(\xi)}{\Pc(\xi)}
\eeq
where we retain the assumption of a perfectly linear PA. Using ${\sf EE}^{\rm linear}(\xi)$, we can obtain the following theorems, which allow us to obtain insights on the structured properties of the practical EE, ${\sf EE}(\xi)$, at least for $\xi\ll1$. The proofs are given in Appendix \ref{Appendix:Theorem1}.

\begin{thm}\label{thm3}
${\sf EE}^{\rm linear}(\xi)$ is a piecewise quasi-concave function over $\xi\geq\zeta \triangleq \left(v+\sqrt{1+v^2}\right)^2/\gamma^2$, where $v=\Pmaxout c_0c_2 / (P_0+\Pmaxout c_0c_1)$. Specifically, ${\sf EE}^{\rm linear}(\xi)$ is quasi-concave over $\zeta\leq \xi\leq1/\way^2$ and also over $1/\way^2< \xi\leq 1$.
\end{thm}

We denote $\xi_{\sf EE}^{\star}$ as the optimal power loading factor that maximizes ${\sf EE}^{\rm linear}(\xi)$, which in general depends on the PA parameters. Typically, $\zeta\approx 0$ as $\gamma\triangleq\Pmaxout/\sigma_z^2$ is large, see e.g., the numerical results in Section~\ref{Sec:NumericalSE}. Assuming $\xi_{\sf EE}^{\star}\geq \zeta$, {\it Theorem~\ref{thm4}} states the solution for $\xi_{\sf EE}^{\star}$.


\begin{thm}\label{thm4}
Assuming $\xi_{\sf EE}^{\star}\geq \zeta$, $\xi_{\sf EE}^{\star}$ equals either $[{\xi}_1^{\star}]^{1/\way^2}_{\zeta}$ or $[{\xi}_2^{\star}]^1_{1/\way^2,}$, where ${\xi}_1^{\star}$ and ${\xi}_2^{\star}$ are the solutions of $\frac{\partial {\sf EE}^{\rm linear}(\xi)}{\partial \xi}=0$ in (\ref{derivative3}) with $(c_1,c_2)$ defined as (\ref{Coeffi1}) and (\ref{Coeffi2}), respectively. Here, notation $[x]^b_a=a$ if $x<a$,  $[x]^b_a=b$ if $x>b$,  and $[x]^b_a=x$ otherwise.\end{thm}

Assuming $\xi_{\sf EE}^{\star}\geq \zeta$, {\it Theorem~\ref{thm4}} shows that there are at most two candidates for $\xi_{\sf EE}^{\star}$. Hence, $\xi_{\sf EE}^{\star}$ can be obtained easily by checking which candidate maximizes ${\sf EE}^{\rm linear}(\xi)$.
Moreover, {\it Proposition~\ref{prop2}} shows that an approximation of $\xi_{\sf EE}^{\star}$ can be obtained in closed form. For the special case of class B PA (i.e., $\way=1$), it follows from {\it Theorem~\ref{thm4}} that the optimal solution is given exactly by $\xi_{\sf EE}^{\star}={\xi}_1^{\star}$, assuming $\xi_{\sf EE}^{\star}\geq \zeta$.

\begin{prop}\label{prop2}
A closed form approximation of $\xi_{\sf EE}^{\star}$, denoted by $\widetilde{\xi}_{\sf EE}^{\star}$, equals either $[\widetilde{\xi}_1^{\star}]^{1/\way^2}_{\zeta}$ or $[\widetilde{\xi}_2^{\star}]^1_{1/\way^2,}$, where $\widetilde{\xi}_i^{\star}$ ($i\in\{1,2\}$) approximates ${\xi}_i^{\star}$ and is given by
\beq\label{XiStar}
{\xi}_i^{\star} \approx \widetilde{\xi}_i^{\star}=\frac{1}{\gamma} \exp\left( { 2+2 {\sf W}\left( \frac{\sqrt{\gamma}}{e v} \right) } \right).
\eeq
\end{prop}
In (\ref{XiStar}), $v$ is defined in Theorem \ref{thm3}, and ${\sf W}(\cdot)>0$ as $\frac{\sqrt{\gamma}}{e v}>0$, so that ${\sf W}(\cdot)$ is unique.

Numerical results in next subsection show the tightness of the EE bound and that $\widetilde{\xi}_{\sf EE}^{\star}$ is a near maximizer of the practical EE, ${\sf EE}(\xi)$.

\subsection{Numerical Results on EE}\label{Sec:NumecialEE}

\begin{figure}[!t]
\psfrag{x}[cc][cc][.8][0]{Power loading factor, $\xi$}
\psfrag{y}[cc][cc][.8][0]{${\sf EE}(\xi),~\MbpJ$}
\psfrag{b}[lc][cc][.7][0]{\sf $\square$ ${\sf EE}^{\rm linear}(\widetilde{\xi}_{\sf EE}^{\star})$;}
\psfrag{c}[lc][cc][.7][0]{\sf $\times$ ${\sf EE}(\widetilde{\xi}_{\sf EE}^{\star})$;}
\psfrag{f}[lc][cc][.7][0]{\sf $\widetilde{\xi}_{\sf EE}^{\star}$ in (\ref{XiStar})}
\psfrag{e                                                       }[lc][cc][.68][0]{\sf ${\sf EE}^{\rm ideal}(\xi)$ in (\ref{EEideal})}
\psfrag{a}[lc][cc][.68][0]{\sf ${\sf EE}^{\rm linear}(\xi)$ in (\ref{EELinear})}
\psfrag{u}[lc][cc][.68][0]{\sf ${\sf EE}(\xi)$ in (\ref{EE}): {\sf class A PA}}
\psfrag{n}[lc][cc][.68][0]{\sf ${\sf EE}(\xi)$ in (\ref{EE}): {\sf class B PA}}
\psfrag{o}[lc][cc][.68][0]{\sf ${\sf EE}(\xi)$ in (\ref{EE}): {\sf $2$-way Doherty PA}}
\begin{center}
\epsfxsize=0.7\textwidth \leavevmode
\epsffile{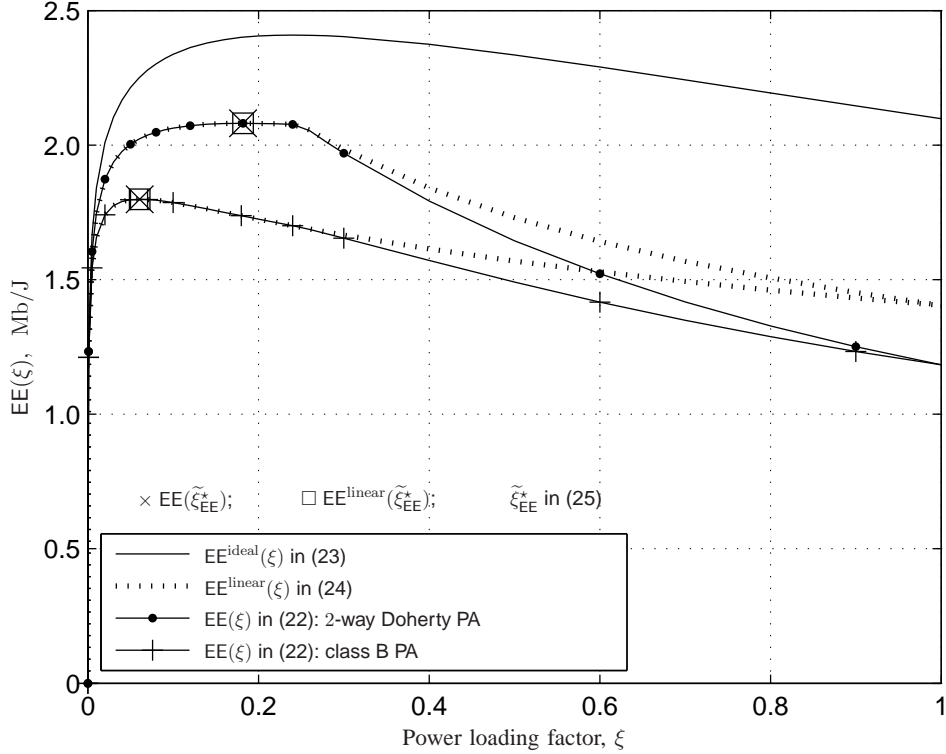}
\caption{Energy efficiency evaluation with $P_{\rm fix}=130\W$ and $c=4.7$.}
\label{Fig:EE}
\end{center}
\end{figure}

To verify the analysis on EE, we evaluate the EE numerically. For power consumption parameters, the macrocell setup in Section \ref{sec:PropPC} is employed. Other parameters are the same as environment given in Section \ref{Sec:NumericalSE}.

Fig. \ref{Fig:EE} shows the EE for class B and $2$-way Doherty PAs. Though the PA specifications, such as the maximum output power and gain, are identical, each of them has different efficiency resulting in different PA parameters in (\ref{Coeffi}). From Fig. \ref{Fig:EE}, we observe that the EE functions are concave ({\it Theorem \ref{thm3}}), and that the Doherty PA achieves the closest EE to the ideal PA's. The EEs, ${\sf EE}(\widetilde{\xi}_{\sf EE}^{\star})$ and ${\sf EE}^{\rm linear}(\widetilde{\xi}_{\sf EE}^{\star})$, are illustrated by `$\times$' and `$\square$,' respectively. As shown in {\it Theorem \ref{thm4}}, $\widetilde{\xi}_{\sf EE}^{\star}$ yields the maximum ${\sf EE}^{\rm linear}(\widetilde{\xi}_{\sf EE}^{\star})$ and it is almost identical to ${\sf EE}(\widetilde{\xi}_{\sf EE}^{\star})$ (which are overlapped in the figure). This is because the practical EE is maximized in the linear region, and the practical SE is also maximized in linear region as shown numerically in the previous section. From the results, we can surmise that the optimal $\widetilde{\xi}_{\sf EE}^{\star}$ in (\ref{XiStar}) is a good approximation of the maximizer of ${\sf EE}(\xi)$.

\section{SE-EE Tradeoff and PA Switching Strategy}\label{Sec:TimeSharing}
\begin{figure}[!t]
\psfrag{x}[cc][cc][.8][0]{${\sf SE}(\xi),~\bpsHz$}
\psfrag{y}[cc][cc][.8][0]{${\sf EE}(\xi),~\MbpJ$}
\psfrag{a                                               }[lc][cc][.68][0]{\sf SE-EE tradeoff with ${\sf PA^{low}}$}
\psfrag{c}[lc][cc][.68][0]{\sf SE-EE tradeoff with ${\sf PA^{high}}$}
\psfrag{v}[lc][cc][1][0]{\sf $\circ$}
\psfrag{z}[lc][cc][1][0]{\sf $\times$}
\psfrag{e}[lc][cc][.68][0]{\sf: (${\sf SE}(\widetilde{\xi}_{\sf SE}^{\star}),{\sf EE}(\widetilde{\xi}_{\sf SE}^{\star})$) from (\ref{XiO})}
\psfrag{n}[lc][cc][.68][0]{\sf: (${\sf SE}(\widetilde{\xi}_{\sf EE}^{\star}),{\sf EE}(\widetilde{\xi}_{\sf EE}^{\star})$) from (\ref{XiStar})}
\psfrag{u}[cc][cc][.68][0]{\sf Fig. \ref{Fig:Sharing}}
\begin{center}
\epsfxsize=0.7\textwidth \leavevmode
\epsffile{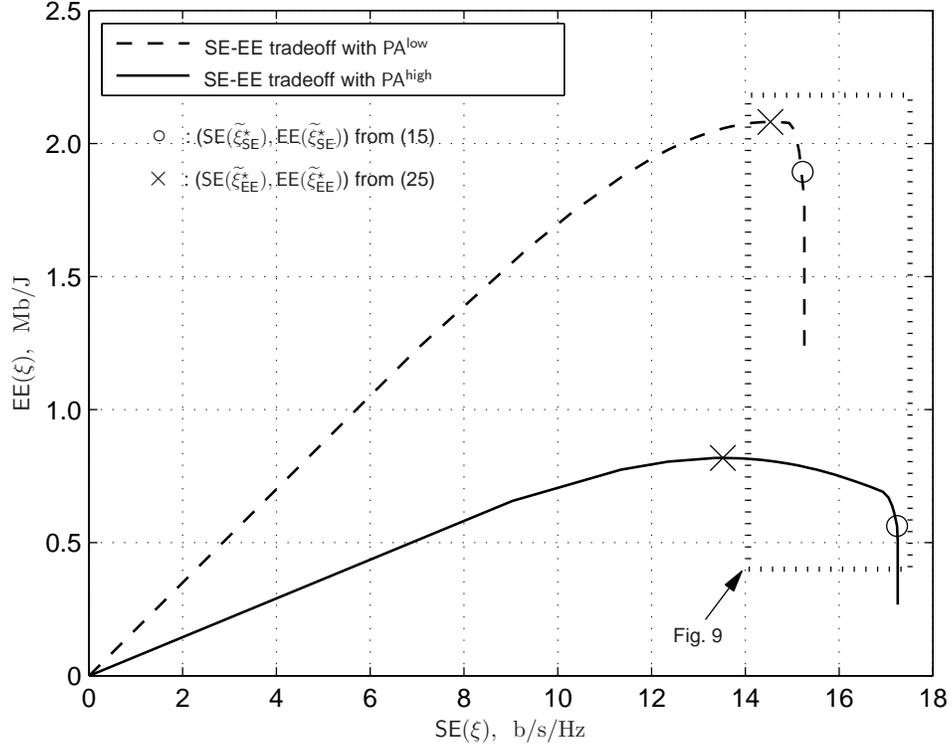}
\caption{SE-EE tradeoff with $2$-way Doherty PAs. ${\sf PA^{low}}$ is a low power PA with $\Pmaxout=25\W$ and $g=55\dB$, and ${\sf PA^{high}}$ is a high power PA with $\Pmaxout=100\W$ and $g=50\dB$.}
\label{Fig:EESE}
\end{center}
\end{figure}

To obtain the practical SE-EE tradeoff, Fig. \ref{Fig:EESE} is regenerated from the results of SE in Fig. \ref{Fig:SE} and EE in Fig. $\ref{Fig:EE}$. In addition to the SE and EE of PA SM2122-44L in subsection \ref{Sec:NumericalSE}, we include the results obtained from a PA SM1720-50 ($\Pmaxout=50\dBm=100\W$ and $g=50\dB$) in Table \ref{Table:DataSheet}. The former and the latter PAs are denoted by ${\sf PA^{low}}$ and ${\sf PA^{high}}$, respectively. We use a $2$-way Doherty PA for all results.

In contrast to the SE-EE tradeoff for an ideal PA which is a decreasing convex function, the EE in the practical SE-EE tradeoff drops rapidly when the SE exceeds beyond a threshold that corresponds to the maximum EE. The closed-form analysis of the SE-EE tradeoff appears intractable. Instead, we focus on the analysis of the tradeoff based on the approximated SE and EE defined (\ref{SEapprox}) and (\ref{EELinear}), respectively.
\begin{prop}\label{prop3}
The Pareto-optimality of the approximated SE-EE tradeoff is characterized as follows:
\begin{description}
\item[i)] For $\minj\{\widetilde{\xi}_{\rm EE}^{\star},\widetilde{\xi}_{\rm SE}^{\star}\}\leq \xi \leq \maxj\{\widetilde{\xi}_{\rm EE}^{\star},\widetilde{\xi}_{\rm SE}^{\star}\}$, the corresponding approximated SE-EE tradeoff is Pareto-optimal: to increase the approximated SE, the approximated EE must decrease, and vice versa.
\item[ii)] For $\xi<\minj\{\widetilde{\xi}_{\rm EE}^{\star},\widetilde{\xi}_{\rm SE}^{\star}\}$, both the approximated SE and EE increase as $\xi$ increases.
\item[iii)] For $\xi>\maxj\{\widetilde{\xi}_{\rm EE}^{\star},\widetilde{\xi}_{\rm SE}^{\star}\}$, both the approximated SE and EE decrease as $\xi$ increases.
\end{description}
\end{prop}
From {\it Proposition \ref{prop3}}, it is sufficient to consider only the region in i), because the remaining regions do not lead to the approximated Pareto-optimal SE-EE tradeoff. In Fig. \ref{Fig:EESE}, the approximated Pareto-optimal SE-EE tradeoff is narrow, which lies between the maximum SE and the maximum EE as indicated by `$\circ$' and `$\times$,' respectively. In cellular communications, however, a wide range of SE-EE tradeoff such as that illustrated by the dotted box in Fig. \ref{Fig:EESE} is desired. This motivates us to use multiple PAs, where one or more PAs are switched on at any time. We call this technique {\em PA switching (PAS)}. Although PAS incurs a switch insertion loss of $G_S$ and an overhead of switching time $\epsilon$ which decrease the SE and EE, we may obtain a better tradeoff of SE-EE from the degree of freedom of choosing different PAs.

For simplicity of description, we consider two PAs, PA-$1$ and PA-$2$; subsequent results are readily extended to multiple PAs. Let the SE and EE of PA-$i$, $i\in\{1,2\}$, including the switch insertion loss $G_S$, be ${\sf SE}_i'(\xi)$ and ${\sf EE}_i'(\xi)=\frac{\bw{\sf SE}_i'(\xi)}{\Pc^i(\xi)}$, respectively, where $\Pc^i(\xi)$ is the total power consumption with PA-$i$. In the following subsections, we apply the PAS technique to two systems, namely, frequency division duplex (FDD) and time division duplex (TDD) systems, and derive their SEs and EEs.


\subsection{PA Switching for FDD Systems}
Consider $\FN$ FDD frames each with length of $T$. For PAS, we assume PA-$1$ is used for the first $\Find$ frames, then PA-$1$ is switched to PA-$2$ which consumes $\epsilon$ seconds, and finally PA-$2$ is used for the remaining $\FN-\Find$ frames. Defining the time sharing factor as $\kappa \triangleq \frac{\Find}{\FN}$, $0\leq \kappa \leq 1$, the achievable SE and EE from PAS can be derived as follows:
\beq \label{SE_sharing_FDD}
\begin{split}
{\sf SE}_{\rm s}^{\rm FDD}(\xi,\kappa)
& = \frac{ \Find T {\sf SE}_1'(\xi)+ (\FN-\Find)T{\sf SE}_2'(\xi)}{\FN T +\epsilon}\\
& = \frac{\FN T}{\FN T +\epsilon}\left( \kappa {\sf SE}_1'(\xi)+(1-\kappa){\sf SE}_2'(\xi)\right)
\end{split}
\eeq
and
\beq\label{EE_sharing_FDD}
\begin{split}
{\sf EE}_{\rm s}^{\rm FDD}(\xi,\kappa)
&=\frac{\FN T \bw{\sf SE}_{\rm s}^{\rm FDD}(\xi,\kappa)}{\Find T \Pc^1(\xi)+(\FN-\Find)T \Pc^2(\xi)}\\
&=\frac{\FN T}{\FN T +\epsilon}
\left(
\frac{{\sf EE}_1'(\xi){\sf EE}_2'(\xi)\left(\kappa {\sf SE}_1'(\xi)+(1-\kappa){\sf SE}_2'(\xi)\right)}{\kappa{\sf SE}_1'(\xi){\sf EE}_2'(\xi) + (1-\kappa){\sf SE}_2'(\xi){\sf EE}_1'(\xi)}
\right),
\end{split}
\eeq
where $\epsilon=0$ if $\kappa=0$ or if $\kappa=1$ (i.e., no switching), and $\epsilon >0$ otherwise. Here, we ignore the switch power consumption as it is relatively negligible compared to $P_{c}^i(\xi)$.

\subsection{PA Switching for TDD Systems}
In TDD systems, the downlink (DL) and uplink (UL) frames are transmitted alternately from BS to UE and from UE to BS. Here, we assume that $\epsilon$ is less than UL frame length which is typically true. For example, one LTE frame consumes a time period of $10\ms$ \cite{LTE}, while the switching time is much less than $1\ms$ (refer to the PA turn-on time in Table \ref{Table:DataSheet} which consumes most of the switching time). We therefore can switch the PAs between consecutive DL frames while receiving UL frame, without switching time overhead. The corresponding SE and EE can be readily obtained from (\ref{SE_sharing_FDD}) and (\ref{EE_sharing_FDD}) by setting $\epsilon$ to be zero. Note that the switching insertion loss is still incurred.

\subsection{Numerical Results and Discussion on PAS}\label{Sec:Results}

\begin{figure}[!t]
\psfrag{a                                            }[lc][cc][.68][0]{\sf $G_S=0\dB$, $\epsilon=0\us$, ideal}
\psfrag{n}[lc][cc][.68][0]{\sf $G_S=1\dB$, $\epsilon=0\us$, TDD}
\psfrag{e}[lc][cc][.68][0]{\sf $G_S=1\dB$, $\epsilon=10\us$, FDD}
\psfrag{c}[lc][cc][.68][0]{\sf $G_S=1\dB$, $\epsilon=1\ms$, FDD}
\psfrag{P}[rc][cc][.68][0]{\sf $+210\%$ EE}
\psfrag{Q}[cc][cc][.68][0]{\sf $-12\%$ SE}
\psfrag{x}[cc][cc][.8][0]{${\sf SE}(\xi),~\bpsHz$}
\psfrag{y}[cc][cc][.8][0]{${\sf EE}(\xi),~\MbpJ$}
\psfrag{A}[cc][cc][.68][0]{${\sf A}$}
\psfrag{B}[cc][cc][.68][0]{${\sf D}$}
\psfrag{C}[cc][cc][.68][0]{${\sf B}$}
\psfrag{D}[cc][cc][.68][0]{${\sf C}$}
\psfrag{E}[cc][cc][.68][0]{${\sf E}$}
\psfrag{z}[lc][cc][.68][0]{SE-EE tradeoff with ${\sf PA^{high}}$}
\begin{center}
\epsfxsize=0.7\textwidth \leavevmode
\epsffile{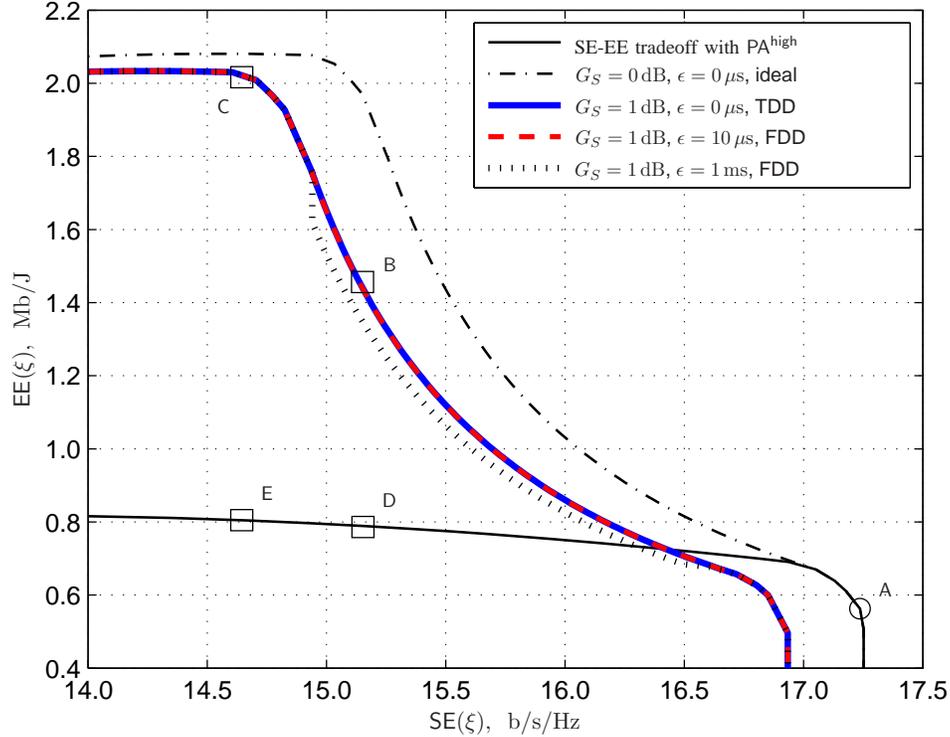}
\caption{SE-EE tradeoff with switching $2$-way Doherty PAs, ${\sf PA^{low}}$ and ${\sf PA^{high}}$, when $\epsilon=10\us$, $L_S=1\dB$, $T=10\ms$, and $\FN=20$.}
\label{Fig:Sharing}
\end{center}
\end{figure}

The PAS is useful for adaptive systems where the traffic and channel conditions change dynamically. Fig. \ref{Fig:Sharing} shows the SE-EE tradeoff with PAS between ${\sf PA^{low}}$ and ${\sf PA^{high}}$. For comparison, we include the results of a single PA ${\sf PA^{high}}$ and an ideal switching, namely, $G_S=0\dB$ and $\epsilon=0$. For practical switching, the switch insertion loss $G_S$ is set to $1\dB$, and $\epsilon$ is set to $0\us$ for TDD frame, while $10\us$ and $1\ms$ are used for FDD frames. We consider $K=20$ frames with $T=10\ms$ for each frame length. From Fig. \ref{Fig:Sharing}, we can verify that an SE-EE tradeoff is substantially improved by PAS. For example, let us consider the TDD system. The EE can be improved by around $210\%$ ($323\%$) if we reduce SE by $12\%$ ($15\%$) from ${\sf A}$ to ${\sf B}$ (${\sf C}$), respectively, as marked in Fig. \ref{Fig:Sharing}. In contrast, if a single PA ${\sf PA^{high}}$ is used instead, the EE is improved by only around $64\%$ ($68\%$) with the same reduction of SE from ${\sf A}$ to ${\sf D}$ (${\sf E}$). Next, consider the FDD system. Even with a switching time that is $10\%$ of the frame size, i.e., $\epsilon=1\ms$, a better SE-EE tradeoff is observed for most of the tradeoff region.


To implement the PAS in practice, the network overhead to obtain full channel state information at the transmitter can be significant, but it can be resolved by limiting the PA numbers with limited feedback information. Other issue is the increased form factor; however, this may not be significant issue in cellular networks where the BSs are already large in form factor due to other circuits. Furthermore, even with a small number of PAs, as we show in our recent work \cite{JoHoSu12WCL}, significant performance gain can be achieved. In the near future, advancement of semiconductor technology will help further reduce the related concerns with form factor and hardware cost, making the proposed PAS an even more convincing technology for any type of transmitters.

\section{Conclusion}
In this paper, we provided a theoretical analysis of the spectral efficiency and energy efficiency (SE-EE) tradeoff of OFDM systems by taking into account the practical non-ideal effects of the power amplifiers (PAs). Optimal power loading factors of PA are derived to achieve the maximum SE and EE. We identified the problem of a narrow SE-EE tradeoff region due to the nonlinearity and inefficiency of the practical PAs, and proposed a PA switching that is a useful technique to achieve a wide SE-EE tradeoff. Future studies include the SE-EE analysis of multiuser communication systems, MIMO systems, and a more accurate PA model with a memory effect and nonlinearity at low power regime.

\appendices

\renewcommand{\theequation}{\thesection.\arabic{equation}}
\setcounter{equation}{0}
\section{}\label{Appendix:Table}
See Table \ref{Table:DataSheet}.

\begin{table*}[t]
\renewcommand{\arraystretch}{0.4}
\centering
\caption{Power Amplifier Characteristics (ascending order of $\Pmaxout$).}\label{Table:DataSheet}
\vspace{-0.5cm}
\subfigure[]{%
\resizebox{0.9\textwidth}{!}{
\begin{tabular}
{|c|c|c|c|c|c|c|c|c|c|}\hline
& & & & & & & & &\\
{PA\#} & {Model} &  {$\Pmaxout(\dBm)$} & {$g(\dB)$} & {$V_{\rm PA}$ (Volt)} & {$C_{\rm PA}$ (mA)} & {$\Pmaxin (\dBm)$}& {Frequency $(\GHz)$} &
{turn-on time ($\us$)} &
{Institution}\\
& & & & & & & & &\\\hline
& & & & & & & & &\\
1	&	MAX2242	&	5	&	28.5	&	3.3	&	50	&	10	&	2.4	{--}	 2.5	&	 1.5	 &	 MAXIM	 \\
2	&	FMPA2151	&	7	&	31	&	3.3	&	280	&	0	&	2.4	{--}	 2.5	&	 {--}	 &	 FAIRCHILD Semiconductor 	\\
3	&	FMPA2151	&	7	&	33	&	3.3	&	375	&	0	&	4.9	{--}	 5.9	&	 {--}	 &	 FAIRCHILD Semiconductor 	\\
4	&	PA1137	&	8	&	17	&	2	&	20	&	10	&	2	{--}	2.2	 &	{--}	 &	 tyco Electronics	 \\
5	&	ADL5570	&	10	&	29	&	3.5	&	100	&	{--}	&	2.3	{--}	2.4	 &	1.0	&	 Analog Devices	 \\
6	&	MAX2242	&	13	&	28.5	&	3.3	&	90	&	10	&	2.4	{--}	 2.5	&	 1.5	 &	 MAXIM	 \\
7	&	MAX2840	&	15	&	22.8	&	3.3	&	155	&	{--}	&	5.15	{--}	 5.35	 &	 1.5	 &	 MAXIM	\\
8	&	BGA6289	&	15	&	13	&	4.1	&	88	&	{--}	&	1.95	{--}	 2.5	&	 {--}	 &	 NXP Semiconductors	\\
9	&	AWT6134	&	16	&	23.5	&	3.4	&	130	&	10	&	1.75	{--}	 1.78	 &	 {--}	 &	 ANADIGICS	\\
10	&	AWT6138	&	16	&	15	&	3.4	&	57	&	10	&	1.85	{--}	 1.91	&	 {--}	 &	 ANADIGICS	 \\
11	&	AWT6252	&	16	&	24.5	&	3.4	&	54	&	10	&	1.92	{--}	 1.98	 &	 {--}	 &	 ANADIGICS	\\
12	&	AWT6252	&	16	&	20.5	&	3.4	&	54	&	10	&	1.92	{--}	 1.98	 &	 {--}	 &	 ANADIGICS	\\
13	&	RF2192	&	16	&	22	&	3.4	&	150	&	10	&	0.824	{--}	 0.849	&	 40	&	 EF MICRO-DEVICES	\\
14	&	RF2196	&	16	&	20	&	3.4	&	160	&	10	&	1.85	{--}	 1.91	&	 {--}	 &	 EF MICRO-DEVICES	\\
15	&	RF3163	&	16	&	24	&	3.4	&	125	&	10	&	0.824	{--}	 0.849	&	 46	&	 EF MICRO-DEVICES	\\
16	&	RF3164	&	16	&	28	&	3.4	&	130	&	10	&	1.85	{--}	 1.91	&	 46	&	 EF MICRO-DEVICES	\\
17	&	RF3165	&	16	&	27	&	3.4	&	130	&	10	&	1.75	{--}	 1.78	&	 46	&	 EF MICRO-DEVICES	\\
18	&	RF6100-1	&	16	&	26	&	3.4	&	135	&	10	&	0.824	{--}	 0.849	&	 46	 &	 EF MICRO-DEVICES	\\
19	&	AP172-317	&	17	&	33	&	3.3	&	140	&	20	&	1.8	{--}	 2.5	&	 {--}	 &	 SKYWORKS	\\
20	&	SKY65006	&	17	&	30	&	3.3	&	110	&	10	&	2.4	{--}	 2.5	&	 {--}	 &	 SKYWORKS	\\
21	&	RMPA5255	&	18	&	33	&	3.3	&	230	&	{--}	&	4.9	{--}	 5.9	&	 1.0	 &	 FAIRCHILD Semiconductor 	\\
22	&	MAX2841	&	18	&	22.5	&	3.3	&	260	&	{--}	&	5.15	{--}	 5.35	 &	 1.5	 &	 MAXIM	\\
23	&	FMPA2151	&	19	&	31	&	3.3	&	600	&	0	&	2.4	{--}	 2.5	&	 {--}	 &	 FAIRCHILD Semiconductor 	\\
24	&	FMPA2151	&	19	&	33	&	3.3	&	600	&	0	&	4.9	{--}	 5.9	&	 {--}	 &	 FAIRCHILD Semiconductor 	\\
25	&	RMPA2458	&	19	&	31.5	&	3.3	&	103	&	5	&	2.4	{--}	 2.5	 &	 1.0	 &	 FAIRCHILD Semiconductor 	\\
26	&	RF5117	&	21	&	26	&	3	&	200	&	10	&	1.8	{--}	2.8	 &	{--}	 &	 EF MICRO-DEVICES	 \\
27	&	RF5189	&	21	&	25	&	3	&	220	&	10	&	2.4	{--}	2.5	 &	{--}	 &	 EF MICRO-DEVICES	 \\
28	&	SST13LP01	&	21	&	34	&	3.3	&	340	&	{--}	&	4.9	{--}	 5.8	&	 0.2	 &	 Silicon Storage Technology, Inc.	\\
29	&	RF5117	&	22	&	26	&	3	&	500	&	10	&	1.8	{--}	2.8	 &	{--}	 &	 EF MICRO-DEVICES	 \\
30	&	RMPA2455	&	22	&	30	&	5	&	195	&	10	&	2.4	{--}	 2.5	&	 1.0	 &	 FAIRCHILD Semiconductor 	\\
31	&	MAX2242	&	22	&	28.5	&	3.3	&	300	&	10	&	2.4	{--}	 2.5	&	 1.5	 &	 MAXIM	 \\
32	&	AP172-317	&	22.5	&	33	&	3.3	&	220	&	20	&	1.8	{--}	 2.5	 &	 {--}	 &	 SKYWORKS	 \\
33	&	MAX2247	&	23	&	29.5	&	3	&	305	&	5	&	2.4	{--}	 2.5	&	 1.5	 &	 MAXIM	 \\
34	&	SST12LP00	&	23	&	27	&	3.3	&	115	&	{--}	&	2.4	{--}	 2.5	&	 {--}	 &	 Silicon Storage Technology, Inc.	\\
35	&	SST12LP14	&	23	&	31	&	3.3	&	290	&	{--}	&	2.4	{--}	 2.5	&	 0.1	 &	 Silicon Storage Technology, Inc.	\\
36	&	SST13LP01	&	23	&	28	&	3.3	&	260	&	{--}	&	2.4	{--}	 2.485	&	 0.1	 &	 Silicon Storage Technology, Inc.	\\
37	&	AP178-321	&	23.5	&	19	&	3.3	&	186	&	20	&	1.8	{--}	 2.5	 &	 {--}	 &	 SKYWORKS	 \\
38	&	MAX2247	&	24	&	29.5	&	3.3	&	307	&	5	&	2.4	{--}	 2.5	&	 1.5	 &	 MAXIM	 \\
39	&	AP172-317	&	24	&	33	&	3.3	&	240	&	20	&	1.8	{--}	 2.5	&	 {--}	 &	 SKYWORKS	\\
40	&	CX65003	&	24.5	&	11.5	&	5	&	138	&	15	&	1.4	{--}	 2.5	 &	 {--}	 &	 SKYWORKS	 \\
41	&	ADL5570	&	25	&	29	&	3.5	&	440	&	{--}	&	2.3	{--}	2.4	 &	1.0	&	 Analog Devices	 \\
42	&	ADL5571	&	25	&	29	&	3.3	&	450	&	{--}	&	2.5	{--}	2.7	 &	1.0	&	 Analog Devices	 \\
43	&	RF2163	&	25	&	19	&	3.3	&	378	&	15	&	1.8	{--}	2.5	 &	{--}	 &	 EF MICRO-DEVICES	 \\
44	&	MAX2247	&	25	&	30.5	&	4.2	&	345	&	5	&	2.4	{--}	 2.5	&	 1.5	 &	 MAXIM	 \\
45	&	SST12LP14	&	25	&	31	&	3.3	&	340	&	{--}	&	2.4	{--}	 2.5	&	 0.1	 &	 Silicon Storage Technology, Inc.	\\
46	&	NE552R479A	&	26	&	11	&	3	&	217	&	19	&	2.45	{--}	 2.45	 &	 {--}	 &	 CEL California Eastern Laboratories	\\
47	&	PA1153	&	26.4	&	28.5	&	15	&	250	&	15	&	1.8	{--}	 2	&	 {--}	 &	 tyco Electronics	\\
48	&	PA1133	&	26.5	&	29	&	15	&	200	&	15	&	1.85	{--}	 1.91	 &	 {--}	 &	 tyco Electronics	\\
49	&	ADL5571	&	27	&	27.5	&	5	&	620	&	{--}	&	2.5	{--}	 2.7	&	 1.0	 &	 Analog Devices	\\
50	&	RF2114	&	27	&	36	&	6.5	&	300	&	12	&	0.001	{--}	 0.6	&	 0.1	 &	 EF MICRO-DEVICES	 \\
51	&	RF2161	&	27	&	30	&	3	&	477	&	6	&	1.85	{--}	 2	&	 {--}	&	 EF MICRO-DEVICES	\\
52	&	RF2186	&	27	&	31	&	3	&	668	&	6	&	1.85	{--}	 2	&	 {--}	&	 EF MICRO-DEVICES	\\
53	&	RF5117	&	27	&	26	&	5	&	500	&	10	&	1.8	{--}	2.8	 &	{--}	 &	 EF MICRO-DEVICES	 \\
54	&	RF5176	&	27	&	26	&	3	&	476	&	6	&	1.85	{--}	 2	&	 {--}	&	 EF MICRO-DEVICES	\\
55	&	AWT6252	&	27.5	&	26.5	&	3.4	&	423	&	10	&	1.92	 {--}	 1.98	&	 {--}	&	 ANADIGICS	\\
56	&	AWT6134	&	28	&	26	&	3.4	&	475	&	10	&	1.75	{--}	 1.78	&	 {--}	 &	 ANADIGICS	 \\
57	&	AWT6134	&	28	&	25	&	3.4	&	462	&	10	&	1.75	{--}	 1.78	&	 {--}	 &	 ANADIGICS	 \\
58	&	AWT6138	&	28	&	26	&	3.4	&	487	&	10	&	1.85	{--}	 1.91	&	 {--}	 &	 ANADIGICS	 \\
59	&	RF3163	&	28	&	28.5	&	3.4	&	455	&	10	&	0.824	{--}	 0.849	&	 46	 &	 EF MICRO-DEVICES	\\
60	&	RF3164	&	28	&	28	&	3.4	&	460	&	10	&	1.85	{--}	 1.91	&	 46	&	 EF MICRO-DEVICES	\\
61	&	RF3165	&	28	&	28	&	3.4	&	460	&	10	&	1.75	{--}	 1.78	&	 46	&	 EF MICRO-DEVICES	\\
62	&	RF6100-1	&	28	&	29	&	3.4	&	465	&	10	&	0.824	{--}	 0.849	&	 46	 &	 EF MICRO-DEVICES	\\
63	&	RF2132	&	28.5	&	29	&	4.8	&	327	&	12	&	0.824	{--}	 0. 849	&	 0.1	 &	 EF MICRO-DEVICES	\\
64	&	RF2146	&	28.5	&	18.5	&	4.8	&	393	&	12	&	1.5	{--}	 2	&	 0.55	 &	 EF MICRO-DEVICES	\\
65	&	RF6100-4	&	28.5	&	28	&	3.4	&	535	&	10	&	1.85	 {--}	 1.91	&	 46	 &	 EF MICRO-DEVICES	\\
66	&	CX65105	&	28.5	&	25	&	5	&	470	&	7	&	1.7	{--}	 2.2	&	 {--}	 &	 SKYWORKS	\\
67	&	SKY65162-70LF	&	28.8	&	20	&	5	&	306	&	{--}	&	 0.869	{--}	 0.96	 &	 {--}	 &	 SKYWORKS	\\
68	&	RF2192	&	29	&	30	&	3	&	715	&	10	&	0.824	{--}	 0.849	&	 40	&	 EF MICRO-DEVICES	\\
69	&	RF2196	&	29	&	27	&	3	&	755	&	10	&	1.85	{--}	 1.91	&	 {--}	 &	 EF MICRO-DEVICES	\\
70	&	MGA-43228	&	29.2	&	38.5	&	5	&	500	&	{--}	&	2.3	 {--}	 2.5	&	 {--}	 &	Avago Technologies	\\
71	&	MGA-43328	&	29.3	&	37.3	&	5	&	470	&	{--}	&	2.5	 {--}	 2.7	&	 {--}	 &	Avago Technologies	\\
72	&	AWT6104M5	&	30	&	30	&	3.5	&	714	&	10	&	1.85	{--}	 1.91	 &	 {--}	 &	 ANADIGICS	\\
73	&	HMC457QS16G/E	&	30.5	&	25	&	5	&	500	&	15	&	2.01	 {--}	 2.17	 &	 {--}	 &	 Hittite Microwave corporation	 \\
74	&	HMC453QS16G/E	&	33	&	8	&	6.5	&	725	&	{--}	&	2.01	 {--}	 2.17	&	 {--}	 &	 Hittite Microwave corporation	\\
75	&	SM0825-33/33H	&	33	&	20	&	12	&	1100	&	4	&	0.8	 {--}	 2.5	&	 {--}	 &	 Stealth Microwave	\\
76	&	SM1025-36DMQ2	&	33	&	10	&	12	&	800	&	{--}	&	1	{--}	 2.5	 &	 {--}	 &	 Stealth Microwave	\\
77	&	SM1025-37MQ2	&	33	&	10	&	12	&	1600	&	{--}	&	1	 {--}	 2.5	&	 {--}	 &	Stealth Microwave	\\
78	&	PA1110	&	33	&	10	&	10	&	725	&	28	&	1.8	{--}	2	 &	{--}	 &	 tyco Electronics	 \\
79	&	PA1132	&	33	&	22	&	12	&	725	&	15	&	1.8	{--}	2	 &	{--}	 &	 tyco Electronics	 \\
80	&	SM1727-34HS	&	34	&	33	&	12	&	1200	&	1	&	1.7	{--}	 2.7	 &	 {--}	 &	 Stealth Microwave	\\
81	&	SM1727-34HSQ	&	34	&	36.5	&	12	&	1200	&	1	&	 1.7	 {--}	 2.7	 &	 {--}	&	 Stealth Microwave	\\
82	&	SM2023-34HS	&	34	&	33	&	12	&	1200	&	1	&	2	{--}	 2.3	 &	 {--}	 &	 Stealth Microwave	\\
83	&	PA1157	&	36	&	24.5	&	10	&	1350	&	15	&	2	{--}	 2.2	 &	 {--}	 &	 tyco Electronics	\\
84	&	PA1159	&	36.2	&	23.5	&	10	&	1700	&	28	&	2.3	 {--}	 2.4	&	 {--}	 &	 tyco Electronics	\\
& & & & & & & & &\\
\hline
\end{tabular}
}}
\end{table*}

\begin{table*}
\renewcommand{\arraystretch}{0.4}
\centering
\subfigure[\it Continued]{%
\resizebox{0.9\textwidth}{!}{
\begin{tabular}{|c|c|c|c|c|c|c|c|c|c|}\hline
& & & & & & & & &\\
{PA\#} & {Model} &  {$\Pmaxout(\dBm)$} & {$g(\dB)$} & {$V_{\rm PA}$ (Volt)} & {$C_{\rm PA}$ (mA)} & {$\Pmaxin (\dBm)$}& {Frequency $(\GHz)$} &
{turn-on time ($\us$)} &
{Institution}\\
& & & & & & & & &\\\hline
& & & & & & & & &\\
85	&	PA1162	&	36.2	&	30	&	10	&	1450	&	11	&	0.8	{--}	 0.96	 &	 {--}	 &	 tyco Electronics	\\
86	&	SM04060-37HS	&	37	&	36	&	12	&	1800	&	1	&	0.4	 {--}	 0. 6	&	 {--}	&	 Stealth Microwave	\\
87	&	SM04093-36HS	&	37	&	34	&	12	&	1600	&	1	&	0.4	 {--}	 0. 925	&	 {--}	&	 Stealth Microwave	\\
88	&	SM5659-37S	&	37	&	20	&	12	&	2300	&	20	&	5.6	{--}	 5.9	 &	 1.0	 &	 Stealth Microwave	\\
89	&	SM5759-37HS	&	37	&	39	&	12	&	2300	&	2	&	5.7	{--}	 5.9	 &	 1.0	 &	 Stealth Microwave	\\
90	&	PA1182	&	37.5	&	23	&	28	&	1000	&	15	&	2.3	{--}	 2.4	 &	 {--}	 &	 tyco Electronics	\\
91	&	PA1223	&	37.5	&	25	&	28	&	1000	&	15	&	2.11	 {--}	 2.17	&	 {--}	&	 tyco Electronics	\\
92	&	PA1224	&	37.5	&	25	&	28	&	1000	&	15	&	2	{--}	 2.2	 &	 {--}	 &	 tyco Electronics	\\
93	&	PA1186	&	38	&	29	&	28	&	1000	&	15	&	0.8	{--}	 0.96	&	 {--}	 &	 tyco Electronics	\\
94	&	XD010-42S-D4F/Y	&	39	&	30	&	28	&	930	&	20	&	0.869	 {--}	 0.894	&	 {--}	&	 Sirenza Micro Devices	\\
95	&	SM0822-39	&	 39	&	 45	&	12	&	3500	 &	 -4	&	0.8	 {--}	 2.2	&	 1.0 &  Stealth Microwave	 \\
96	&	SM0825-40Q	&	40	&	39	&	12	&	5500	&	1	&	0.8	{--}	 2.5	 &	 {--}	 &	 Stealth Microwave	\\
97	&	SM2023-41	&	41	&	55	&	12	&	4500	&	-13	&	2	{--}	 2.3	 &	 {--}	 &	 Stealth Microwave	\\
98	&	SM2027-41LS	&	41	&	51	&	12	&	6000	&	-7	&	2	{--}	 2.7	 &	 {--}	 &	 Stealth Microwave	\\
99	&	SM4450-41L	&	41	&	55	&	12	&	5000	&	-13	&	4.4	{--}	 5	&	 1.0	 &	 Stealth Microwave	\\
100	&	SM1822-42LS	&	42	&	52	&	12	&	5500	&	-8	&	1.8	{--}	 2.2	 &	 {--}	 &	 Stealth Microwave	\\
101	&	SM3338-43	&	43	&	50	&	12	&	8500	&	-6	&	3.3	{--}	 3.8	 &	 1.0	 &	 Stealth Microwave	\\
102	&	SM5053-43L	&	43	&	55	&	12	&	9200	&	-7	&	5	{--}	 5.3	 &	 1.0	 &	 Stealth Microwave	\\
103	&	SM7785-43A	&	43	&	48	&	12	&	9500	&	{--}	&	7.725	 {--}	 8.5	 &	 {--}	 &	Stealth Microwave	\\
104	&	SM1923-44L	&	44	&	55	&	12	&	8200	&	-8	&	1.9	{--}	 2.3	 &	 {--}	 &	 Stealth Microwave	\\
105	&	SM2025-44L	&	44	&	55	&	12	&	8500	&	-10	&	2	{--}	 2.5	 &	 {--}	 &	 Stealth Microwave	\\
\rowcolor[gray]{.7}{\textbf{106}}	&	{\textbf{SM2122-44L}	}&	 {\textbf{44}}	&	 {\textbf{55}}	&	 {\textbf{12}}	&	{\textbf{8200}}	 &	 {\textbf{-9}}	&	 {\textbf{2.1}	 {--}	\textbf{ 2.2}}	&	 & { \textbf{Stealth Microwave}}	 \\
107	&	SM2325-44	&	44	&	55	&	12	&	8000	&	-10	&	2.3	{--}	 2.5	 &	 {--}	 &	 Stealth Microwave	\\
108	&	SM2025-46L	&	46.3	&	52	&	12	&	15000	&	-7	&	2	 {--}	 2.5	&	 {--}	 &	 Stealth Microwave	\\
109	&	SM04548-47L	&	47	&	55	&	12	&	14000	&	-8	&	0.45	 {--}	 0.48	&	 {--}	&	 Stealth Microwave	\\
110	&	SM2023-47L	&	47	&	55	&	12	&	15000	&	-7	&	2	{--}	 2.3	 &	 {--}	 &	 Stealth Microwave	\\
111	&	SM3134-47L	&	47	&	55	&	12	&	15000	&	-6	&	3.1	{--}	 3.4	 &	 {--}	 &	 Stealth Microwave	\\
112	&	SM3436-47L	&	47	&	56	&	12	&	15000	&	-6	&	3.4	{--}	 3.6	 &	 {--}	 &	 Stealth Microwave	\\
\rowcolor[gray]{.7} {\textbf{113}}	&	{\textbf{SM1720-50}}	&	{\textbf{50}}	&	 {\textbf{50}}	&	 {\textbf{12}}	&	{\textbf{27000}}	 &	 {\textbf{2}}	&	 {\textbf{1.7	 {--}	 2}}	&	 {\textbf{{--}}}	 &	 {\textbf{Stealth Microwave}}	 \\
113	&	SM2325-50L	&	50	&	59	&	12	&	31000	&	-9	&	2.3	{--}	 2.5	 &	 {--}	 &	 Stealth Microwave	\\
115	&	SM1819-52LD	&	52	&	45	&	30	&	11000	&	{--}	&	1.8	{--}	 1.9	 &	 {--}	 &	 Stealth Microwave	\\
& & & & & & & & &\\
\hline
\end{tabular}
}}
\end{table*}

\renewcommand{\theequation}{\thesection.\arabic{equation}}
\setcounter{equation}{0}
\section{Derivation for (\ref{fy13})}\label{Appendix:Eq}
{\it Joint pdf $f_{Y}(y,S=0)$}: Given $S=0$, i.e., $\ampA \leq \amax$, we have $W=\fPA(\ampA)e^{j\theta} = \sqrt{g} \ampA e^{j\theta}$. From assumption $A1$, the conditional pdf of $W$ is a truncated complex Gaussian as
\beq\label{rvW}
{{
\begin{split}
&f_{W}\left(w|S=0\right) =
\left\{
\begin{array}{ll}
\frac{1}{\Pr(S=0)}\frac{1}{\pi  \gPin} \exp\left(-\frac{|w|^2}{\gPin}\right), & |w| < \bmax\\
0, & |w| \geq \bmax.
\end{array}
\right.
\end{split}
}}
\eeq
The pdf of the AWGN $\awgN$ is
\beq\label{rvN}
{{
\begin{array}{c}
f_{\awgN}(\awgn|S=0) = f_{\awgN}(\awgn|S=1) = f_{\awgN}(\awgn)  = \frac{1}{\pi \sigma_z^2} \exp\left(-\frac{|\awgn|^2}{\sigma_z^2}\right).
\end{array}
}}
\eeq
From \eqref{flatfading}, we can express the joint pdf as follows:
\beq
{{
\begin{split}
f_{\y}(y,S=0)
&=
\Pr(S=0)\left(f_{W}\left(w|S=0\right)\conv f_{\awgN}(\awgn|S=0)\right)\\
&=
\Pr(S=0)\int_{\tau\in\mathbb{C}} f_{W}\left(\tau|S=0\right) f_\awgN\left(y-\tau\right)  d\tau  \\
&=
\Pr(S=0)\int_{0}^{\infty} \int_{0}^{2\pi} f_{W}\left(\ampr e^{j\phi}\big|S=0\right)  f_\awgN\left(y-\ampr e^{j\phi}\right) \left|J(\phi,\ampr)\right| d\phi d \ampr \label{fy}
\end{split}
}}
\eeq
where `$\conv$' is the linear convolution operator and $|J(\phi,\ampr)|={\ampr}$ is the Jacobian \cite{PaPi02book}. Noting $(f_{W}|S=0)=0$ if $|w|\geq \bmax$, and using (\ref{rvW}) and (\ref{rvN}) to (\ref{fy}), we further derive
\beq\nonumber
{{
\begin{split}
f_{Y}(y,S=0)
&=
\frac{1}{ \pi^2 \gsigmaxsquare \sigma_z^2} \int_{0}^{\bmax}
\int_{0}^{2\pi} {\ampr}\exp\!\left(-\frac{{\ampr}^2}{\gsigmaxsquare}-\frac{|y-{\ampr} e^{j\phi}|^2}{\sigma_z^2}\right) d\phi d {\ampr} \\
&=
\frac{1}{\pi^2 \gsigmaxsquare \sigma_z^2}\int_{0}^{\bmax}
{\ampr} \exp\left(-\frac{{\ampr}^2}{\gsigmaxsquare}-\frac{{\ampr}^2+|y|^2}{\sigma_z^2}\right)\int_{0}^{2\pi} \exp\left(\frac{2{\ampr} |y| \cos (\theta-\phi)}{\sigma_z^2}\right) d\phi d {\ampr} \\
&=
\frac{2}{ \pi \gsigmaxsquare \sigma_z^2}\int_{0}^{\bmax} {\ampr}
\exp\left(-\frac{{\ampr}^2}{\gsigmaxsquare}-\frac{{\ampr}^2+|y|^2}{\sigma_z^2}\right)
{\sf I}_0\left(\frac{2{\ampr}|y|}{\sigma_z^2}\right) d {\ampr}\\
&=
\frac{1}{\pi\left(\gsigmaxsquare +\sigma_z^2\right)}\exp\left(-\frac{|y|^2}{\gsigmaxsquare+\sigma_z^2}\right)\int_{0}^{\rhomax}\left[\frac{1}{2}\exp\left(-\frac{\mu(y)+\rho}{2}\right){\sf I}_{0}\left(\sqrt{\mu(y) \rho}\right) \right] d \rho
\end{split}
}}
\eeq
where $\theta$ denotes the angle of $y$ and ${\sf I}_0(\cdot)$ is a modified Bessel function of first kind. The last equality is obtained after some mathematical manipulations, for which we observe the function outside the integral is a pdf of a normal distribution with a zero mean and variance $(\gsigmaxsquare+\sigma_z^2)$ and the function inside the integral is a pdf of a noncentral chi-squared random variable with one degree of freedom. Since the cdf of a noncentral chi-squared random variable is obtained as $1-{\sf Q}_1(\sqrt{\mu(y)},\sqrt{\rhomax})$, we can readily arrive at (\ref{fy1}).

{\it Joint pdf $f_{Y}(y,S=1)$}: Given $S=1$, i.e., $\ampA > \amax$, we have $W=\fPA(\ampA)e^{j\theta} =\sqrt{g}\amax e^{j\theta} =\bmax e^{j\theta}$. Thus, the amplitude is a constant while the phase  $\theta$ is uniformly distributed. Therefore, we can express the conditional pdf of $W$ given $S=1$ as
\beq\label{fWS1}
{{
f_{W}\left(w|S=1\right) = c  \delta\left(|w|-\bmax\right)
}}
\eeq
where $\delta(\cdot)$ is the Dirac delta function and $c= ({2\pi\bmax})^{-1}$ is a constant obtained from the normalization $\int_{w\in\mathbb{C}}f_{W}\left(w|S=1\right) d w = 1$. Then the joint pdf is similarly to give
\beq
{{
\begin{split}
f_{\y}(y,S=1)
=
\Pr(S=1)\int_{0}^{\infty} \!\!\int_{0}^{2\pi} \ampr f_{W}\left({\ampr} e^{j\phi}\big|S=1\right) f_{\awgN}\left(y-{\ampr} e^{j\phi}\right) d\phi d {\ampr}.
\label{fy3a}
\end{split}
}}
\eeq
Substituting (\ref{fWS1}) and $f_{\awgN}(\awgn|S=1)=f_{\awgN}(z)$ in (\ref{rvN}) into (\ref{fy3a}), and integration over ${\ampr}$, we get
\beq
{{
\begin{split}
f_{\y}(y,S=1)
&=
\frac{\Pr(S=1)}{2\pi^2\sigma_z^2} \int_{0}^{2\pi} \exp\left(-\frac{\left|y-\bmax e^{j\phi}\right|^2}{\sigma_z^2}\right) d\phi\nonumber\\
&=
\frac{\Pr(S=1)}{2\pi^2\sigma_z^2}\exp\left(-\frac{|y|^2+\bmax^2}{\sigma_z^2}\right)2\pi{\sf I}_0\left(\frac{2\bmax|y|^2}{\sigma_z^2}\right)\nonumber\\
&=
\frac{1}{\pi\sigma_z^2}\exp\left(-\frac{|y-\bmax|^2}{\sigma_z^2}\right) \left[\Pr(S=1)\exp\left(\frac{-2\bmax \yreal}{\sigma_z^2}\right) {\sf I}_0\left(\frac{2\bmax|y|}{\sigma_z^2}\right)\right]\nonumber
\end{split}
}}
\eeq
where the function outside $[\cdot]$ is a pdf of a normal distribution $\mathcal{CN}(\bmax,\sigma_z^2)$. Thus, we get (\ref{fy3}).


\setcounter{equation}{0}
\section{Proof of (\ref{MI_multipath4})}\label{Appendix:Multi}

Henceforth, we take the indices in the subscript to be modulo $N$, e.g., $X_{N+i}=X_{(N+i)\bmod{N}}=X_i$.
We express the mutual information as
\beq
{{
\begin{split}
I(\Xvec ;\Yvec)=I(\xvec ;\yvec)\overset{(a)}{=}& \sum_{t=0}^{N-1}I(\x_t;\y_0,\cdots,\y_{N-1}| \x_0, \cdots, \x_{t-1})\\
\overset{(b)}{\geq}& \sum_{t=0}^{N-1}\!I(\x_t;\y_0,\cdots,\y_{t+L-1}| \x_0, \cdots, \x_{t-1})
\label{MI_multipath}
\end{split}
}}
\eeq
where (a) follows from the chain rule of mutual information and (b) follows from the data processing inequality (by discarding the received signals $\y_{t+L},\cdots,\y_{N}$). The inequality in (b) is typically tight from numerical experiments, because $\x_t$ is only present in $\y_{t},\cdots,\y_{t+L-1}$; therefore, intuitively the discarded signals do not directly contribute to the information on $\x_t$ (although they contribute to the information on the interfering terms in $\y_{t},\cdots,\y_{t+L-1}$). The summand in \eqref{MI_multipath} for $t\geq L$ can be lower bounded as follows:
\beq
{{
\begin{split}
I(\x_t;\y_0,\cdots,\y_{t+L-1}| \x_0, \cdots, \x_{t-1})
&\overset{(a)}{=}I(\x_t;\y_{t},\cdots,\y_{t+L-1}| \x_0, \cdots, \x_{t-1})\\
&\overset{(b)}{=}I(\x_t; Y_t', \cdots, Y_{t+L-1}'| \x_0, \cdots, \x_{t-1}) \\
&\overset{(c)}{=}I(\x_t; Y_t', \cdots, Y_{t+L-1}')  \\
&\overset{(d)}{\geq}I(\x_t; Y_t'')\triangleq I^{\text{LB}}_t
\label{MI_multipath2}
\end{split}
}}
\eeq
where (a) follows from the independence of  $\x_t$ and $\{\y_{1},\cdots, \y_{t-1}\}$ given $\{\x_0, \cdots, \x_{t-1}\}$; (b) follows from the definition $\y'_{t+i}=\y_t-\sum_{j=i+1}^{L-1} h_j\x_{t+i-j} $ for $i=0,\cdots,L-1$; (c) follows from the fact that $\y'_{t}, \cdots, \y'_{t+L-1}$ consist only of the signal terms $X_t,\cdots,X_{N-1}$ and noise; and (d) follows from the data processing inequality where $Y_t''$ is the maximum ration combining (MRC) of $Y_t', \cdots, Y_{t+L-1}'$.
For additive independent interference with variance $\sigma^2$, the mutual information is lower bounded by the channel where the noise is treated as AWGN with the same variance $\sigma^2$ \cite{CoTh06book}. Hence, a lower bound of $I(\x_t; Y_t'')$, denoted by $I^{\text{LB}}_t$, is given by the mutual information of the following channel:
\beqn\label{MI_multipath3}
\y'''_t &=& \; \fPA(\ampA_t) e^{\theta_t} + Z'_t
\eeqn
where $Z'_{t}\sim\mathcal{CN}(0,\sigma_z^2/|h'_t|^2)$ and $h'_t$ is derived by
\beq\label{MRCch}
h'_t =\sqrt{\frac{\gPin|h_0|^2}{\sigma_z^2} + \sum_{i=1}^{L-1} \frac{\gPin|h_i|^2}{ \sigma_z^2+\gPin \sum_{j=0}^{i-1}|h_j|^2} }.
\eeq
The channel in (\ref{MRCch}) is the equivalent channel after MRC of the current and $(L-1)$ future received signals, where we take yet-to-be decoded, transmitted signals as interferences. The channel in (\ref{MI_multipath3}) is the flat fading channel \eqref{flatfading} with equivalent noise variance given by the original noise variance divided by the equivalent channel gain. Hence, we can obtain $I^{\text{LB}}_t$ from \eqref{MI} directly. In summary, from (\ref{MI_multipath}) and (\ref{MI_multipath2}), the mutual information is lower bounded by (\ref{MI_multipath4}).

\setcounter{equation}{0}
\section{Proof of Theorems and Propositions}\label{Appendix:Theorem1}
\begin{IEEEproof}[Proof of Theorem \ref{thm1}] After substituting (\ref{EntropyApp}) into (\ref{SEapprox}), we can derive the second derivative of ${\sf SE}^{\rm IBO}(\xi)$ with respect to $\xi$ as follows:
\beq\nonumber
{{
\begin{split}
\frac{\partial^2 {\sf SE}^{\rm IBO}(\xi)}{\partial^2 \xi}& = 
\frac{-1}{\ln 2}\Bigg( \frac{\gamma^2}{(1+\gamma\xi)^2}+ e^{-\xi^{-1}}\xi^{-4}+ e^{-\xi^{-1}}\xi^{-3}\left(\xi^{-1}-2\right)\left(\ln
\frac{1}{\pi \sigma_z^2}-\frac{1}{\xi}\right)\Bigg)
\end{split}
}}
\eeq
${\sf SE}^{\rm IBO}(\xi)$ is concave over $\xi$ if $\max\left(0,\frac{-1}{\ln(\pi\sigma_z^2)}\right) \leq \xi \leq \frac{1}{2}$ as the second derivative is negative.\end{IEEEproof}

\begin{IEEEproof}[Proof of Theorem \ref{thm2}] Since ${\sf SE}^{\rm IBO}(\xi)$ is concave assuming $\max\left(0,\frac{-1}{\ln(\pi\sigma_z^2)}\right) \leq \xi \leq \frac{1}{2}$ from {\it Theorem \ref{thm1}}, we can find the maximizer of ${\sf SE}^{\rm IBO}(\xi)$ by solving $\frac{\partial {\sf SE}^{\rm IBO}(\xi)}{\partial \xi}=0$, which gives (\ref{C1Eq}).\end{IEEEproof}

\begin{IEEEproof}[Proof of Proposition \ref{prop1}] Since $\gamma\xi\gg 1$ with practical value of $\gamma\triangleq\Pmaxout/\sigma_z^2$ and typical value of $\xi$, we can approximate $1+\gamma\xi\approx \gamma\xi$ in (\ref{C1Eq}), and get $
\xi e^{\frac{1}{\xi}}
 =
-\frac{1}{\xi} +1 - \ln\left(\pi e\sigma_z^2\right)\approx - \ln\left(\pi e\sigma_z^2\right)
$ where we discard relatively small terms to obtain the closed form solution $\widetilde{\xi}_{\sf SE}^{\star}$ in (\ref{XiO}). \end{IEEEproof}

\begin{IEEEproof}[Proof of Theorem \ref{thm3}]
First, assume $(c_1,c_2)$ is fixed over all $\xi$ (the dependence according to \eqref{Coeffi} will be considered shortly).
The first derivative of EE with respect to $\xi$ in (\ref{EELinear}) is given by
\beq\label{derivative3}
{{
\begin{split}
&\frac{\partial {\sf EE}^{\rm linear}(\xi)}{\partial \xi} =\frac{\bw}{2\sqrt{\xi}\left(v_1+v_2\sqrt{\xi}\right)^2}\left(\frac{2}{\ln 2}\frac{\gamma}{1+\gamma\xi} \left( v_1\sqrt{\xi}+v_2 \xi\right) -v_2 \log_2(1+\gamma\xi) \right)
\end{split}
}}
\eeq
where $v_1=P_0+\Pmaxout c_0c_1$, and $v_2=\Pmaxout c_0c_2$.
Clearly $v_2 \log_2(1+\gamma\xi)$ is  increasing in $\xi$. It can be shown that $\frac{\gamma}{1+\gamma\xi} \left( v_1\sqrt{\xi}+v_2 \xi\right)$ is decreasing in $\xi$ if $\xi\geq \zeta$, where $\zeta$ is defined in Theorem~\ref{thm3}. Assuming $\xi\geq\zeta$, ${\sf EE}^{\rm linear}(\xi)$ then has at most one turning point and thus ${\sf EE}^{\rm linear}(\xi)$ is a quasi-concave function; it cannot be a quasi-convex function because ${\sf EE}^{\rm linear}(\xi)$ is decreasing in $\xi$ for sufficiently large $\xi$. Now note that $(c_1,c_2)$ is in fact constant for $0\leq\xi \leq \frac{1}{\way^2}$, and also constant for $\frac{1}{\way^2} < \xi \leq 1$. Thus, ${\sf EE}^{\rm linear}(\xi)$ is a quasi-concave function for $\zeta \leq \xi \leq \frac{1}{\way^2}$, and also for $\frac{1}{\way^2} <\xi \leq 1$.
\end{IEEEproof}

\begin{IEEEproof}[Proof of Theorem \ref{thm4}] First, consider the optimal solution that maximizes ${\sf EE}^{\rm linear}(\xi)$ over $\zeta\leq \xi \leq \frac{1}{\way^2}$.
From~{\it Theorem \ref{thm3}}, ${\sf EE}^{\rm linear}(\xi)$ is a quasi-concave function. Thus the optimal solution is given by the turning point ${\xi}_1^{\star}$. In other words, the solution of $\frac{\partial {\sf EE}^{\rm linear}(\xi)}{\partial \xi}=0$ with $(c_1,c_2)$ defined as (\ref{Coeffi1}), if ${\xi}_1^{\star}$ lies between $\zeta$ and $\frac{1}{\way^2}$.
Since ${\xi}_1^{\star}$ is a turning point, the optimal solution must be $\zeta$ if ${\xi}_1^{\star}<\zeta$, while the optimal solution must be $\frac{1}{\way^2}$ if ${\xi}_1^{\star}>\frac{1}{\way^2}$. More concisely, the optimal solution is $[{\xi}_1^{\star}]^{1/\way^2}_{\zeta}$. Similarly, the optimal solution that maximizes ${\sf EE}^{\rm linear}(\xi)$ over $\frac{1}{\way^2}\leq \xi \leq 1$ can be shown to be $[{\xi}_2^{\star}]^1_{1/\way^2}$.\end{IEEEproof}

\begin{IEEEproof}[Proof of Proposition \ref{prop2}] The same approximation used in the proof of {\it Theorem 2}, i.e., $1+\gamma\xi\approx \gamma\xi$, gives an equality $\frac{2}{\ln 2}\frac{1}{\xi} \left( v_1\sqrt{\xi}+v_2 \xi\right) =v_2 \log_2(\gamma\xi)$ to make (\ref{derivative3}) be a zero. After some mathematical manipulations of the equality, we can find the unique solution given by (\ref{XiStar}).
\end{IEEEproof}

\begin{IEEEproof}[Proof of Proposition \ref{prop3}] Suppose $\widetilde{\xi}_{\rm EE}^{\star}\leq \widetilde{\xi}_{\rm SE}^{\star}$. From {\it Theorems} \ref{thm1} and \ref{thm3} and {\it Propositions} \ref{prop1} and \ref{prop2}, the approximated SE and EE are concave functions over $\xi$, and $\widetilde{\xi}_{\rm SE}^{\star}$ and $\widetilde{\xi}_{\rm EE}^{\star}$ are their maximizers, respectively. Thus, we can show the following. i) For $\xi\geq\widetilde{\xi}_{\rm EE}^{\star}$, ${\sf EE}(\xi)$ decreases as $\xi$ increases. For $\xi\leq\widetilde{\xi}_{\rm SE}^{\star}$, ${\sf SE}(\xi)$ increases as $\xi$ increases. Thus, for $\widetilde{\xi}_{\rm EE}^{\star}\leq \xi \leq \widetilde{\xi}_{\rm SE}^{\star}$, ${\sf EE}(\xi)$ decreases, while ${\sf SE}(\xi)$ increases as $\xi$ increases. ii) For $\xi<\widetilde{\xi}_{\rm EE}^{\star}$, both ${\sf EE}(\xi)$ and ${\sf SE}(\xi)$ increase as $\xi$ increase. iii) For $\xi>\widetilde{\xi}_{\rm SE}^{\star}$, both ${\sf EE}(\xi)$ and ${\sf SE}(\xi)$ decrease as $\xi$ increases. The proof is completed by showing the similar analysis for the case when $\widetilde{\xi}_{\rm EE}^{\star} < \widetilde{\xi}_{\rm SE}^{\star}$.
\end{IEEEproof}


\end{document}